\documentclass[submission,Phys]{SciPost}

\usepackage[utf8]{inputenc} % input encodings (allow UTF-8 input)
\usepackage[T1]{fontenc} 	% selecting font encodings (use 8-bit T1 fonts)
\usepackage[english]{babel} % multilingual support (English language/hyphenation)

\usepackage[bitstream-charter]{mathdesign}
\usepackage{geometry} 		% flexible and complete interface to document dimensions
\usepackage{amsmath} 		% math package (American Mathematical Society)
\usepackage{amsthm} 		% math package (typesetting theorems using AMS style)
\usepackage{mathtools} 		% math package (fixes various deficiencies of amsmath)
\usepackage{float} 			% floating objects such as figures and tables
\usepackage{graphicx} 		% enhanced support for graphics
\usepackage{tabularx} 		% tabulars with adjustable-width columns
\usepackage{booktabs} 		% professional-quality tables
\usepackage{color, xcolor} 	% foreground and background colour management
\usepackage{pdfpages} 		% inclusion of external multi-page PDF documents
\usepackage{extarrows} 		% extra arrows beyond those provided in amsmath
\usepackage{multirow} 		% create tabular cells spanning multiple rows
\usepackage{multicol} 		% intermix single and multiple columns
\usepackage{caption} 		% customising captions in floating environments
\usepackage{subcaption} 	% support for sub-captions (captions for subfigures)
\usepackage{enumitem} 		% control layout of itemize, enumerate, description
\usepackage{setspace} 		% set space between lines (single, onehalf, double)
\usepackage{xspace} 		% define commands that appear not to eat spaces
\usepackage{ragged2e} 		% alternative versions of ragged-type commands
\usepackage{stackrel} 		% enhancement to the \stackrel command
\usepackage{tikz} 			% create PostScript and PDF graphics
\usepackage{braket} 		% Dirac bra-ket notation
\usepackage{bm} 			% access bold symbols in maths mode
\usepackage{tensor} 		% typeset tensors (tensor-style super- and subscripts)
\usepackage{slashed} 		% slash through characters (Feynman slash notation)
\usepackage{siunitx} 		% SI units package (typesetting values with units)
\usepackage{lastpage} 		% reference last page
\usepackage{cite} 			% improved citation handling
\usepackage[normalem]{ulem} % package for underlining
\usepackage{fontawesome} 	% access to web-related icons
\usepackage{tocloft} 		% control over the typography of the TOC, LOF, and LOT
\usepackage{titlesec} 		% interface to sectioning commands (various title styles)
\usepackage{doi} 			% create correct hyperlinks for DOI numbers
\usepackage{hyperref} 		% hypertext links (handel cross-referencing commands)
\usepackage[most]{tcolorbox} 					% coloured and framed text boxes
\usepackage[nameinlink, capitalize]{cleveref} 	% intelligent cross-referencing
\usepackage[nottoc, notlot, notlof]{tocbibind} 	% add references/index/contents to TOC
\usepackage[ruled, vlined]{algorithm2e} 		% floating algorithm environment
\usepackage{makecell}

\hypersetup{
	pdftitle={NAE for jets},
	pdfauthor={Dillon et al.},
	colorlinks=true, 			% false: boxed links, true: colored links
	linkcolor={red!50!black}, 	% color of internal links (sections, pages, etc.)
	citecolor={blue!50!black}, 	% color of citation links (links to bibliography)
	urlcolor={blue!80!black} 	% color of URL links (external links)
} 
\DeclareSymbolFont{usualmathcal}{OMS}{cmsy}{m}{n}
\DeclareSymbolFontAlphabet{\mathcal}{usualmathcal}

\SetArgSty{textnormal}
\SetKwComment{Comment}{{\small\#}~}{}
\SetCommentSty{mycommfont}

\setitemize{itemsep=0pt, parsep=0pt} 				% adjust itemize environment
\setenumerate{itemsep=0pt, parsep=0pt} 				% adjust enumerate environment
\setitemize{itemsep=2pt,topsep=2pt,parsep=0pt,partopsep=0pt,leftmargin=*}
\setenumerate{itemsep=0pt,topsep=2pt,parsep=0pt,partopsep=0pt,labelindent=3pt,leftmargin=*}
\newcommand{\pd}{p_\text{data}}

\newcommand{\Langle}{\bigl\langle}
\newcommand{\Rangle}{\bigr\rangle}
\newcommand{\XLangle}{\Bigl\langle}
\newcommand{\XRangle}{\Bigr\rangle}

\newcommand\one{\leavevmode\hbox{\small1\normalsize\kern-.33em1}}
\newcommand{\loss}{\mathcal{L}} 	% loss value

\newcommand{\arXiv}[2][]{%
	\ifthenelse{\equal{#1}{}}%
	{\href{http://arxiv.org/abs/#2}{arXiv:#2}}%
	{\href{http://arxiv.org/abs/#2}{arXiv:#2~[#1]}}}

\newcommand{\gev}{\text{GeV}}

\def\slashchar#1{\setbox0=\hbox{$#1$}           % set a box for #1
   \dimen0=\wd0                                 % and get its size
   \setbox1=\hbox{/} \dimen1=\wd1               % get size of /
   \ifdim\dimen0>\dimen1                        % #1 is bigger
      \rlap{\hbox to \dimen0{\hfil/\hfil}}      % so center / in box
      #1                                        % and print #1
   \else                                        % / is bigger
      \rlap{\hbox to \dimen1{\hfil$#1$\hfil}}   % so center #1
      /                                         % and print /
   \fi}

\graphicspath{{./plots/final_plots/}}                % define graphics path

\begin{document}

\begin{center}{\Large \textbf{
A Normalized Autoencoder for LHC Triggers
}}\end{center}

\begin{center}
  Barry M. Dillon\textsuperscript{1},
  Luigi Favaro\textsuperscript{1},
  Tilman Plehn\textsuperscript{1},
  Peter Sorrenson\textsuperscript{2}, and
  Michael Kr\"amer\textsuperscript{3}
\end{center}

\begin{center}
  {\bf 1} Institut f\"ur Theoretische Physik, Universit\"at Heidelberg, Germany\\
  {\bf 2} Heidelberg Collaboratory for Image Processing, Universit\"at Heidelberg, Germany \\
  {\bf 3} Institute for Theoretical Particle Physics and Cosmology (TTK), RWTH Aachen University, Germany \\
\end{center}

\begin{center}
\today
\end{center}

\section*{Abstract}
         {\bf Autoencoders are an effective analysis tool for the LHC, as
           they represent one of its main goal of finding physics beyond the
           Standard Model. The key challenge is that
           out-of-distribution anomaly searches based on the compressibility 
           of features do not apply to the
           LHC, while existing density-based searches lack performance. We
           present the first autoencoder which identifies anomalous
           jets symmetrically in the directions of higher and lower
           complexity. The normalized autoencoder combines a standard
           bottleneck architecture with a well-defined probabilistic
           description. It works better than all available
           autoencoders for top vs QCD jets and reliably identifies
           different dark-jet signals.}

         \begin{center}
	   \includegraphics[width=0.65\textwidth]{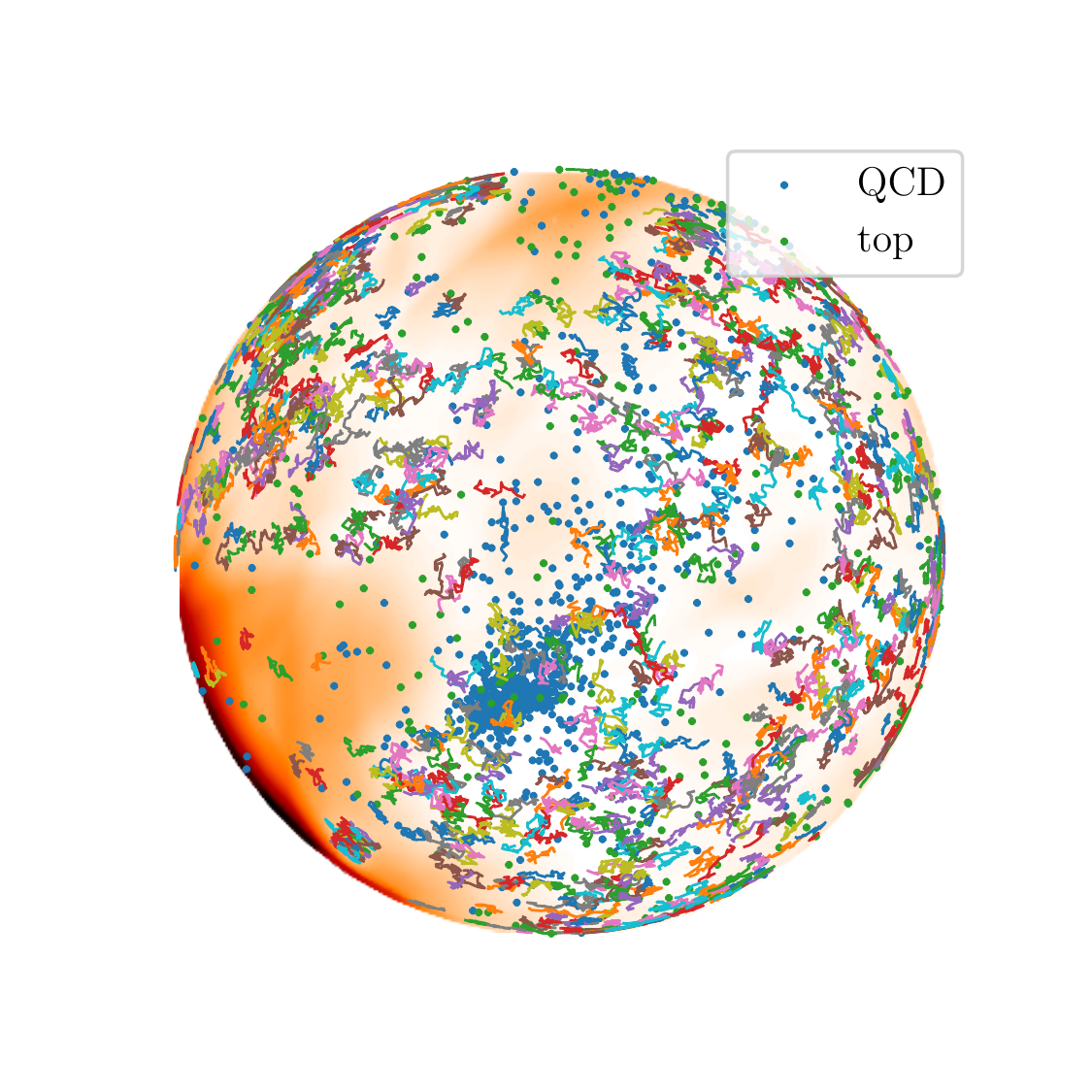}
         \end{center}

         \clearpage
%\vspace*{50pt}
\noindent\rule{\textwidth}{1pt}
\tableofcontents\thispagestyle{fancy}
\noindent\rule{\textwidth}{1pt}
%\vspace{10pt}

\clearpage
%%%%%%%%%%%%%%%%%%%%%%%%%%%%%%%%%%%%%%%%%%%%%%%%%%
\section{Introduction}
\label{sec:intro}

The big goal of the LHC is to discover physics beyond the Standard
Model (BSM) and to identify new properties of the fundamental
constituents of matter. Until now, we pursue BSM searches based on
pre-defined theory hypotheses. The upcoming LHC runs need to
supplement targeted searches with (i) analyses of phase space regions
linked to an effective theory extension of the Standard Model and (ii)
searches for anomalous effects defined as not explained by the
Standard Model. Both of these strategies share the ambitious goal of
understanding all aspects of LHC data using fundamental physics ---
with the secret goal of bringing down the Standard Model.\medskip

A key concept in the search for anomalies in LHC data is that as few as possible
assumptions should be made about the potential signals in the data.
The techniques used for this are strongly influenced by developments
in modern machine learning (ML).  Autoencoders (AEs) are simple tools
for anomaly searches, based on a bottleneck in the mapping of a data
representation onto itself.  They can, for instance, identify
anomalous jets in a QCD jet
sample~\cite{Heimel:2018mkt,Farina:2018fyg}. Because the bottleneck as
the latent space does not have a well-defined structure, the anomaly
score has to be related to the quality of the reconstruction.

Adding a latent space structure leads us to more complex tools, for example variational autoencoders
(VAEs)~\cite{kingma2014autoencoding}.  In the encoding step for a VAE we map a
high-dimensional data representation to a low-dimensional latent
distribution; the decoding step then generates new high-dimensional
objects. The latent space will encode structures which might not be
apparent in the high-dimensional input data.  Such VAEs work for
anomaly searches with LHC jets~\cite{Cerri:2018anq,Cheng:2020dal}, and
we can trade the reconstruction loss for an anomaly score defined in
latent space. In this case, we benefit from a network architecture
which constructs an optimized latent space, for instance the Dirichlet
VAE~\cite{Dillon:2021nxw}, leading to a mode separation between
background and signal.\medskip

Motivated by their initial success, ML-methods for anomaly detection
at the LHC were developed for anomalous jets~\cite{
  Aguilar-Saavedra:2017rzt, Roy:2019jae, Dillon:2019cqt,
  Aguilar-Saavedra:2020uhm,Atkinson:2021nlt,Kahn:2021drv,Dolan:2021pml,Canelli:2021aps,Buss:2022lxw,Atkinson:2022uzb},
anomalous events~\cite{DAgnolo:2018cun, Hajer:2018kqm,
  Komiske:2019fks, Blance:2019ibf, Romao:2019dvs, DAgnolo:2019vbw,
  Andreassen:2020nkr, Amram:2020ykb, Matchev:2020wwx, Komiske:2020qhg,
  Romao:2020ojy, Dillon:2020quc, Romao:2020ocr, Khosa:2020qrz,
  Mikuni:2020qds,
  vanBeekveld:2020txa,Jawahar:2021vyu,Mikuni:2021nwn,Bradshaw:2022qev},
or to enhance search strategies~\cite{Collins:2018epr,
  DeSimone:2018efk, Collins:2019jip, Mullin:2019mmh, 1815227,
  Park:2020pak,Hallin:2021wme,Barron:2021btf,Finke:2022lsu}.  They
include a first ATLAS analysis~\cite{Aad:2020cws}, experimental
validation~\cite{Komiske:2019jim, Knapp:2020dde}, quantum machine
learning~\cite{Ngairangbam:2021yma}, self-supervised
learning~\cite{Dillon:2021gag,Dillon:2022tmm}, applications to
heavy-ion collisions~\cite{Thaprasop:2020mzp}, the DarkMachines
community challenge~\cite{Aarrestad:2021oeb}, and the
LHC~Olympics~2020 community
challenge~\cite{Kasieczka:2021xcg,Bortolato:2021zic}.

%%%MOD
The problem with these studies is that it is not clear what the
anomalous property of jets or events actually means. The step from AEs
to VAEs implies a change in the way we define anomalous jets. For an
unstructured AE we search for out-of-distribution jets based on the
compressibility of their features. The lack of symmetry in tagging
anomalous top jets vs anomalous QCD jets in the respective other
sample casts doubts on this definition for the LHC. While massive top
jets stick out among low-mass QCD jets, we cannot expect a QCD
jet to be special in a sample of top jets~\cite{Finke:2021sdf} since its simpler 
features will be interpolated by the neural network. A
better-suited definition is based on low-probability regions in the
background phase space
distributions~\cite{Nachman:2020lpy,Batson:2021agz,Dorigo:2021iyy,Caron:2021wmq,Fraser:2021lxm}. Here
we can encode the phase space density, for instance, using cluster
algorithms, VAEs, or a normalizing flow~\cite{Buss:2022lxw}, but none
of these methods are especially successful at identifying anomalous
jets once the signal becomes more challenging than top jets.\medskip

In this paper we present the normalised autoencoder
(NAE)~\cite{yoon2021autoencoding} as a remedy for these issues. It
relies on detecting outliers through the reconstruction loss, but via
an energy-based model~\cite{convnet,du2019implicitEBM}. 
%%%MOD
%One of the goals of this work is to develop an autoencoder which
%can reliably identify anomalous jets, and which is smaller enough to 
%run on an LHC trigger.
%Although the jets we study in the paper would pass trigger selection cuts already,
%the results still demonstrate the utility of this technique on a trigger.
%%%%
%%
One of the goals of this work is to develop an autoencoder which is a robust anomalous jets tagger.
We explore the concept of using autoencoders as of triggers, i.e. tools that can extract
interesting events from a given background with as little bias as possible.
Although the jets we study in the paper would pass trigger selection cuts already,
the results still demonstrate how our approach limits the assumptions
made on BSM signals to data preprocessing rather than latent space structure,
in favor of a more model-agnostic network architecture.
The NAE achieves reliable anomaly detection results without
increasing the size of the network, with the additional components 
affecting only how the model is trained.
The training focuses on minimizing the negative
log-likelihood of the data given the network parameters, but
evaluating a probability distribution.  This means that NAE is a
probabilistic model which samples from the model distribution and
penalizes modes absent from the training data.  For phase space
regions with such modes the NAE training adjusts the energy as the
underlying structure of the latent space, such that the autoencoder
gets a robust OOD detector. Two Langevin Markov Chains in
the latent and phase spaces probe the poor-reconstruction regions and
penalize them. Combining this with the minimization of the training
reconstruction error, we define a minmax loss function that converges
when the training and model distributions match each other.

In Sec.~\ref{sec:net} we first introduce energy-based models as an
alternative to reconstruction losses like an MSE or a likelihood
ratio.  We then describe the NAE setup\footnote[2]{The code is available at \url{https://github.com/heidelberg-hepml/normalized-autoencoders}}, with its efficient way of
sampling the background data manifold in phase space and latent
space. In Sec.~\ref{sec:top} we apply the NAE to the top tagging
dataset~\cite{Butter:2017cot,Kasieczka:2019dbj,Benato:2021olt} and
show that, for the first time, the NAE identifies anomalous top jets
and anomalous QCD jets symmetrically and with high efficiency. Next,
we target two challenging dark jet signals~\cite{Buss:2022lxw} and
confirm the excellent performance of the NAE and its relative
independence of the jet image preprocessing in Sec.~\ref{sec:dark}. In
the Appendix we provide additional details about the NAE and our
implementation.

%%%%%%%%%%%%%%%%%%%%%%%%%%%%%%%%%%%%%%%%%%%%%%%%%%
\section{Network and dataset}
\label{sec:net}

The normalized autoencoder~\cite{yoon2021autoencoding} we will use for
this study is an energy-based modification of a standard AE, as
applied in Ref.~\cite{Heimel:2018mkt}. We will first introduce
energy-based models, mention their challenges, and then describe the
way the NAE modifies the AE training. As input data we use jet images
with standard preprocessing.

%%%%%%%%%%%%%%%%%%%%%%%%%%%%%%%%%%%%%%%%%%%%%%%%%%
\subsection{Energy-based networks}
\label{sec:net_energy}

Energy-Based Models (EBMs) are a class of probability density
estimation models appealing for their flexibility. They are defined
through a normalizable energy function, which is minimized during
training.  This energy function can be chosen as any non-linear
function mapping a point to a scalar value~\cite{song2021howtoEBM},
\begin{align}
  E_\theta(x):\;  \mathbb{R}^D \rightarrow \mathbb{R} \; ,
  \label{eq:energy}
\end{align}
where $D$ is the dimensionality of the phase space. The EBM uses this
energy function to define a probabilistic loss, assuming a Boltzmann
or Gibbs distribution as its probability density over phase space,
\begin{align}
p_\theta(x) = \frac{e^{-E_\theta(x)}}{Z_\theta}
\qquad \text{with} \qquad 
Z_\theta = \int_x dx e^{-E_\theta(x)} \; ,
\label{eq:ebm}
\end{align}
with the partition function $Z_\theta$. We omit an explicit
normalization of the energy by a temperature or some other constant in
this formula.  The main feature of a Boltzmann distribution is that
low-energy states have the highest probability. The EBM loss is the
negative logarithmic probability evaluated as a likelihood over the
model parameters,
\begin{align}
\loss(x)
&= - \log  p_\theta(x) 
= E_\theta(x) + \log Z_\theta 
\qquad \Rightarrow \qquad
\loss = \Langle E_\theta(x) + \log Z_\theta \Rangle_{x \sim \pd} \; ,
\label{eq:ebm_loss}
\end{align}
where we define the total loss as the expectation over the per-sample
loss.  The difference to typical likelihood losses is that the
second, normalization term is unknown.

To train the network we want to minimize the loss in
Eq.\eqref{eq:ebm_loss}, so we have to compute its gradient,
\begin{align}
\nabla_\theta \loss(x) 
= - \nabla_\theta \log p_\theta(x) 
&= \nabla_\theta E_\theta(x)
+ \nabla_\theta \log Z_\theta \notag \\
  &=  \nabla_\theta E_\theta(x)
  + \frac{1}{Z_\theta} \nabla_\theta \int_x dx e^{-E_\theta(x)} \notag \\
  &=  \nabla_\theta E_\theta(x)
  -  \int_x dx \frac{e^{-E_\theta(x)}}{Z_\theta} \nabla_\theta E_\theta(x) \notag \\
&=  \nabla_\theta E_\theta(x)
-  \XLangle \nabla_\theta E_\theta(x) \XRangle_{x \sim p_\theta} \; .
\label{eq:dec}
\end{align}
The first term in this expression can be obtained using automatic
differentiation from the training sample, while the second term is
intractable and must be approximated.  Computing the expectation value
over $\pd(x)$ allows us to rewrite the gradient of the loss as the
difference of two energy gradients
\begin{align}
\Langle \nabla_\theta \loss(x) \Rangle_{x \sim \pd}
= \XLangle - \nabla_\theta \log p_\theta(x)\XRangle_{x \sim \pd}
=  \XLangle \nabla_\theta E_\theta(x) \XRangle_{x \sim \pd}
- \XLangle \nabla_\theta E_\theta(x) \XRangle_{x \sim p_\theta} \; .
\label{eq:diff}
\end{align}
The first term samples from the training data, the second from the
model.  According to the sign of the energy in the loss function, the
contribution from the training dataset is referred to as positive
energy and the contribution from the model as negative energy. One way
to look at the second term is as a normalization which ensures that
$\loss = 0$ for $p_\theta (x) = \pd(x)$. Another way is to view it as
inducing a structure for the minimization of the likelihood.\medskip

One practical way of sampling from $p_\theta(x)$ is to use
Markov-Chain Monte Carlo (MCMC).  We use Langevin Markov Chains, where
the steps are defined by drifting a random walk towards high
probability points according to
\begin{align}
  x_{t+1} = x_t + \lambda_x \nabla_x \log p_\theta(x) + \sigma_x \epsilon_t
  \qquad \text{with} \qquad
  \epsilon_t &\sim \mathcal{N}_{0,1} \; .
\label{eq:lmc}
\end{align}
Here, $\lambda$ is the step size and $\sigma$ the noise standard
deviation. When $2\lambda = \sigma^2$ the equation resembles Brownian
motion and gives exact samples from $p_\theta(x)$ in the limit of
$t\rightarrow +\infty$ and $\sigma \rightarrow 0$.

For ML applications working on images, the high dimensionality of the
data makes it difficult to cover the entire physics space $x$ with
Markov chains of reasonable length. For this reason, it is common to
use shorter chains and to choose $\lambda$ and $\sigma$ to place more
weight on the gradient term than on the noise term. If $2\lambda \neq
\sigma^2$, this is equivalent to sampling from the distribution at a
different temperature defined as
\begin{align}
  T = \frac{\sigma^2}{2\lambda} \; .
\end{align}
There the two parameters are defined in Eq.\eqref{eq:lmc}. By
upweighting $\lambda$ or downweighting $\sigma$ we are effectively
sampling from the distribution at a low temperature, thereby
converging more quickly to the modes of the distribution.

By inspecting the expectation value of the loss in the form of
Eq.\eqref{eq:diff}, we can identify the training as a minmax problem,
where we minimize the energy of the training samples and maximize the
energy of the MCMC samples. This means that the energy of training
data points is pushed downwards.  At the same time the energy of
Markov chains sampled from the energy model distribution will be
pushed upwards.  For instance, if $p_\theta(x)$ reproduces $\pd(x)$
over most of the  phase space $x$, but $p_\theta(x)$ includes an
additional mode, its  phase space region will be assigned large values
of $E_\theta(x)$ through the minimization of the loss. This way, all
modes present in the energy model distribution but missing from the
training distribution $\pd$ will be suppressed.  This process of
adjusting the energy continues until the model reaches the equilibrium
in which the model distribution is identical to the training data
distribution.

Despite the well-defined algorithm, training EBMs is difficult due to
instabilities arising from (i) the minmax optimization, with similar
dynamics to balancing a generator and discriminator in a GAN; (ii)
potentially biased sampling from the MCMC due to a low effective
temperature; and (iii) instabilities in the LMC chains.  Altogether,
stabilizing the training during its different phases requires serious
effort.

%%%%%%%%%%%%%%%%%%%%%%%%%%%%%%%%%%%%%%%%%%%%%%%%%%
\subsection{Normalized autoencoder}
\label{sec:net_nae}

An AE is a two-module function that maps an input to its
reconstruction using an encoder-decoder structure,
\begin{align}
  f_\theta(x): \quad
  \mathbb{R}^D \longrightarrow \mathbb{R}^{D_z} \stackrel{f_D}{\longrightarrow} \mathbb{R}^D \; .
\end{align}
The training minimizes the per-pixel difference between the original
input and its mapping.  A typical choice for the loss function is the
Mean Squared Error (MSE) of this reconstruction. With this definition
a plain AE is not a probabilistic model, since a small reconstruction
error does not correspond to a large likelihood.  However, we can
upgrade the AE to a probabilistic NAE by using the MSE as the energy
function in Eq.\eqref{eq:energy}
\begin{align}
  E_\theta(x)
  = \text{MSE}
  \equiv \frac{1}{N} \sum_\text{pixels} \left| x - f_\theta(x) \right|^2 \; .
\end{align}
This way we can train a probabilistic AE using Eqs.\eqref{eq:diff}
and~\eqref{eq:lmc}.

%%%MOD
%By using the reconstruction error as the energy, the model will learn
%to poorly reconstruct inputs not in the training distribution, as it
%adapts its distribution to match that of the training data. The
%advantage of this is that it guarantees the behavior of the model in
%this region of phase space, especially in the region close to but not
%in the training data distribution. We cannot give such a guarantee for
%a standard autoencoder, which only sees the training distribution and
%could assign arbitrary reconstruction scores to data outside this
%distribution.  In fact this is the source of the problems with standard 
%autoencoders outlined in the introduction, which are solved using the NAE.\medskip
%%%%
%%%%

The NAE training includes two steps. First, we pre-train the baseline
AE with the standard MSE loss, similar to
Ref.~\cite{Heimel:2018mkt}. After the AE pre-training we switch to the NAE loss given in
Eq.\eqref{eq:ebm_loss}. All NAE parameters are given in
Tab.~\ref{tab:mc}.  In the spirit of a proper anomaly search we use
the same network and hyper-parameters throughout this paper. They
reflect a trade-off between sampling quality, training stability, and
tagging performance.

The pre-training phase builds an approximate density estimator exclusively based on 
the training data by minimizing the reconstruction error. Then, the NAE loss 
explores the regions with low energy and guarantees the behavior of the model 
especially in the region close to but not in the training data distribution. 
Here, the mismatch between the data and the model distribution is 
corrected by the inter-play between the two components of the loss function.
We cannot give such a guarantee for a standard autoencoder, which only sees the training distribution and
could assign arbitrary reconstruction scores to data outside this
distribution.  In fact this is the source of the problems with standard 
autoencoders outlined in the introduction, which are solved using the NAE.\medskip

%%%%
As mentioned above, training EBMs is a practical challenge.  Several
algorithms have been proposed to train these networks.  Two well-known
methods based on MCMC samples are Contrastive Divergence (CD) and
Persistent CD.  CD and PCD differ in how they define the initial sample.  
CD uses a sample taken from the data
distribution $\pd(x)$ while PCD samples from a replay buffer made up
of the final state of Markov chains from previous steps of the
optimization. However, these methods are susceptible to creating
spurious high density modes and struggle with full space
coverage~\cite{yoon2021autoencoding}.

We follow a different approach, using the fact that we can train a
regular autoencoder before starting the NAE training. If we accept
that different initializations of the MCMC defined in
Eq.\eqref{eq:lmc} lead to different results, we can tune $\lambda_x$
and $\sigma_x$ in such a way that we can use a sizeable number of
short, non-overlapping Markov
chains~\cite{du2019implicitEBM,grathwohl2019your,nijkamp2019learning}.
Specifically, the proposed algorithm for an efficient training of NAEs
is On-Manifold Initialization (OMI)~\cite{yoon2021autoencoding}.  This
approach is motivated by the observation that sampling the full data
space is inefficient due to its high dimensionality, but the training
data lies close to a low-dimensional manifold embedded in the data
space $x$. All we need to do is to sample close to this
manifold. Since we are using an autoencoder this manifold is defined
implicitly as the image of the decoder network, meaning that any point
in the latent space $z$ passed through the decoder will lie on the
manifold. This means we can first focus on the manifold by taking
samples from a suitably defined distribution in the low-dimensional
latent space, and then map these samples into data space via the
decoder. After that, we perform a series of MCMC steps in the full
ambient data space to allow the Markov chains to minimize the loss
around the manifold.

To sample from the model we first need to define a suitable latent
probability density, which we do as
\begin{align}
  q_\theta(z) &= \frac{e^{-H_\theta(z)}}{\Psi_\theta}
  \qquad \text{with} \qquad
  H_\theta(z) = E_\theta(f_D(z)) \; ,
\label{eq:omi}
\end{align} 
where $H_\theta(z)$ is the latent energy, and $f_D$ is the decoder
network, all in complete analogy to Eq.\eqref{eq:ebm}. Having defined
these quantities, the latent space chain is run as
\begin{align}
  z_{t+1} = z_t + \lambda_z \nabla_z \log q_\theta(z) + \sigma_z \epsilon_t
    \qquad \text{with} \qquad \epsilon_t \sim \mathcal{N}_{0,1} \; .
\label{eq:omi_lmc}
\end{align}
Once we reach a high-density point on the decoder manifold, the final
sample is obtained by running a second input chain according to
Eq.\eqref{eq:lmc}.

During the OMI it is crucial that we cover the entire latent space,
thus a compact structure is preferable. To achieve that, we normalize
the latent vectors so that they lie on the surface of a hypersphere
$\mathbb{S}^{D_z-1}$, allowing for a uniform sampling of the initial
batch in the latent space.  The step size and the noise of both chains
are tuned to give $T<1$. Even if a lower temperature introduces a bias
towards modes with lower reconstruction error, this helps stabilize
the training and obtain finer samples from the MCMC.  Long LMC chains
are affected by instability by two reasons: sudden changes in the
gradients between steps, and diverging energy for both positive and
negative samples due to the loss function being independent of
constant shifts.  Even if these issues are still not well understood
in the ML community, different possible solutions have been
proposed~\cite{du2019implicitEBM} and applied in this work: (i)
clipping gradients in each step; (ii) spectral normalization; (iii) L2
weight normalization; and (iv) L2 normalization on positive and
negative samples.  To decrease the size of both chains, a replay
buffer has been utilized which saves the final points of each latent
chain. In the next iteration the initial sample is either drawn from the
buffer or drawn uniformly from the hypersphere, with the probability of being 
drawn from the buffer being 0.95. 
%and uniformly from the hypersphere instead. 
Finally, an acceptance Metropolis step and a noise annealing
step can be applied.\medskip

The encoder has 5 convolutional
layers. Each layer has 8 filters, except for the last layer with one
filter. The output is then flattened, and two dense layers downsize
the network to the latent space size. The decoder mimics the encoder
with 2 dense layers followed by 4 convolutional layers. All
intermediate activation functions are PReLU. The output activation for
the encoder and the decoder are linear and sigmoid, respectively. For
the latent space dimension we use $D_z = 3$, which is not optimized
for performance, but allows us to visualize the latent space easily.
We run the pre-training for 300 epochs, using
Adam~\cite{Kingma:2014vow} with default parameters.
Additional information on the network architecture can be 
found in \ref{sec:network}

%----------------------------------------------------------
\begin{table}[b!]
\centering
\begin{tabular}{l|@{\hskip 0.3in}c@{\hskip 0.3in}c}
\toprule
LMC parameters          & latent   & input    \\ \midrule
 $\lambda$  & 100           & 50            \\ 
 $\sigma$      & $10^{-2}$         & $10^{-4}$         \\ 
\# of steps & 30 & 30 \\
metropolis    &$\checkmark$          &$\checkmark$          \\ 
annealing     &$        -          $          &$\checkmark$          \\ \midrule
training parameters & pre-AE & NAE \\ \midrule
learning rate &  $10^{-3}$ & $10^{-5}$ \\
iterations & 15k &  40k \\
batch size & 2048 & 128 \\ \bottomrule
\end{tabular}
\caption{LMC and training parameters. The temperature is
  implicitly fixed by the noise and the step size as $T_x = 10^{-7}$
  and $T_z \approx 10^{-6}$.}
\label{tab:mc}
\end{table}
%----------------------------------------------------------

%%%%%%%%%%%%%%%%%%%%%%%%%%%%%%%%%%%%%%%%%%%%%%%%%%
\subsection{Jet Images}
\label{sec:net_jets}

%----------------------------------------------------------
\begin{figure}[t]
  \includegraphics[width=\textwidth]{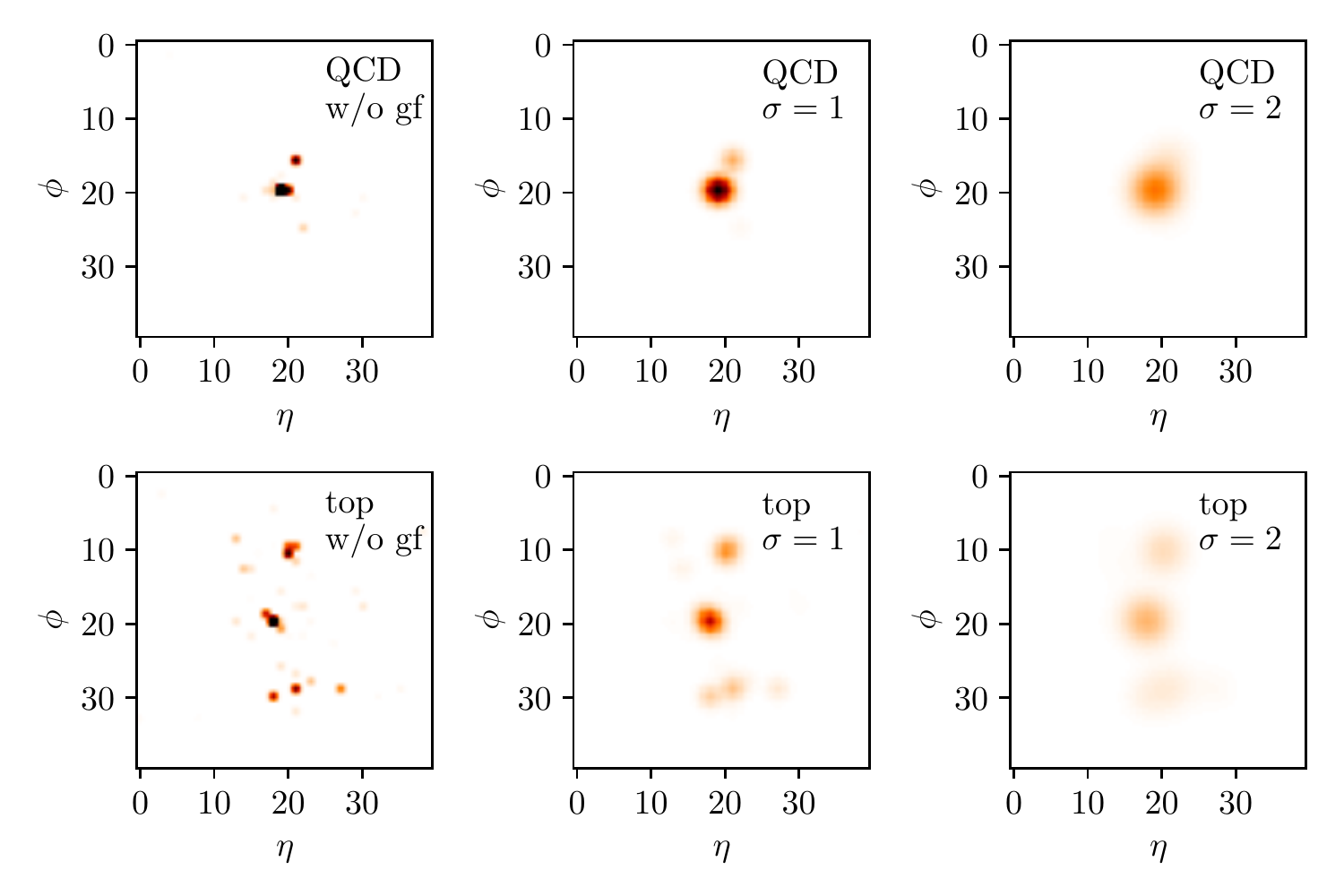}
  \caption{Example QCD and top images without Gaussian filter,
    $\sigma_\text{G}=1$, and $\sigma_\text{G}=2$.}
  \label{fig:gf}
\end{figure}
%----------------------------------------------------------

As our application we choose jet images. The first test will be top vs
QCD, discussed in Sec.~\ref{sec:top}, where we use the top-tagging
dataset~\cite{Butter:2017cot,Kasieczka:2019dbj,Benato:2021olt}, also
used for the AE in Ref.~\cite{Heimel:2018mkt} and the Dirichlet VAE in
Ref.~\cite{Dillon:2021nxw}. We start with anti-$k_T$
jets~\cite{anti-kt} with $R=0.8$, defined by
\textsc{FastJet3.1.3}~\cite{fastjet} as substructure containers. The
top and QCD jets are are required to have
\begin{align} 
  p_{T} = 550~...~650~\gev
  \qquad \text{and} \qquad
  |\eta| < 2 \; .
  \label{eq:topjets}
\end{align}
Before defining the jet image, the jets are pre-processed by centering
each jet around the $k_T$-weighted centroid of all constituents. Then,
the jets are rotated such that their principal axis points to 12
o'clock, and flipped so that the highest $p_T$ region is in the
lower-left quadrant. Then, the constituents are pixelized in $40
\times 40$ images with pixel size $[\Delta \eta, \Delta \phi] = [
  0.029, 0.035]$. The intensity of the pixels is defined by the sum of
$p_T$s within that cell, and finally, the whole image is rescaled by
the total $p_T$ of the event.  To reduce the sparsity we apply a
Gaussian filter to each image. The effect of two Gaussian filters is
illustrated in Fig.~\ref{fig:gf}.  Our top tagging dataset consists of
140k jets for each class, of which we use 100k jets for training, and
the remaining 40k for testing. The jet images are pre-processed with a
Gaussian filter with $\sigma_\text{G}=1$.\medskip

Our second reference dataset, presented in Sec.~\ref{sec:dark} are two
dark-matter-inspired signal samples~\cite{Buss:2022lxw}.  The
underlying model is hidden valleys, with a light and strongly
interacting dark
sector~\cite{Strassler:2006im,Morrissey:2009tf,Knapen:2021eip}.
Particles produced in this dark sector can decay within that sector
and form a dark shower. Such a dark shower will eventually switch to
SM-fragmentation and form either a semi-visible
jet~\cite{Cohen:2015toa,Cohen:2017pzm,Pierce:2017taw,Beauchesne:2017yhh,
  Bernreuther:2019pfb,Bernreuther:2020vhm} or a pure, modified QCD
jet~\cite{Heimel:2018mkt}.  We will refer to the semi-visible jets as
the Aachen dataset and the modified QCD jets as the Heidelberg
dataset. Compared to the QCD background, the Aachen dataset is mostly
more sparse, whereas the Heidelberg dataset includes an additional
decay structure.  The two signal samples and the QCD background sample
are generated just like to the top jet sample, but with
\begin{align}
  p_{T,j} = 150~...~300~\gev
  \qquad \text{and} \qquad
  |\eta_j| < 2 \; .
\label{eq:darkjets}
\end{align}
As for the top jets, a Gaussian filter improves the network training
and we use a filter with a $\sigma_\text{G} = 1$, but some additional
precautions are needed for dark jets. We know that for an efficient
identification of both of the dark jets we need to reweight the jet
images. Unlike in Ref.~\cite{Buss:2022lxw} we now apply the same
pixel-wise remapping for both dark jet signals, namely
\begin{align}
  p_T \quad \to p_T^n
  \qquad \text{with} \qquad
  n = 0.01, 0.1, 0.2, 0.3, 0.5 \; .
\label{eq:remap}
\end{align}
The goal is to reduce the dependence of the autoencoder performance on
this remapping for different signals.

%%%%%%%%%%%%%%%%%%%%%%%%%%%%%%%%%%%%%%%%%%%%%%%%%%
\section{QCD vs top jets}
\label{sec:top}

%----------------------------------------------------------
\begin{figure}[b!]
  \includegraphics[width=0.495\textwidth]{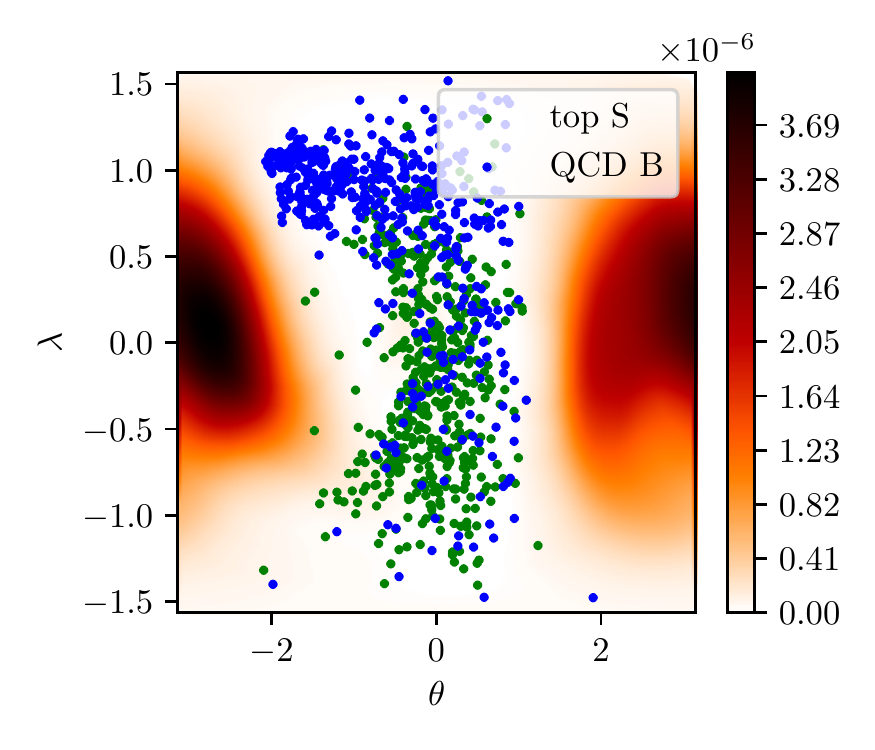}
  \includegraphics[width=0.495\textwidth]{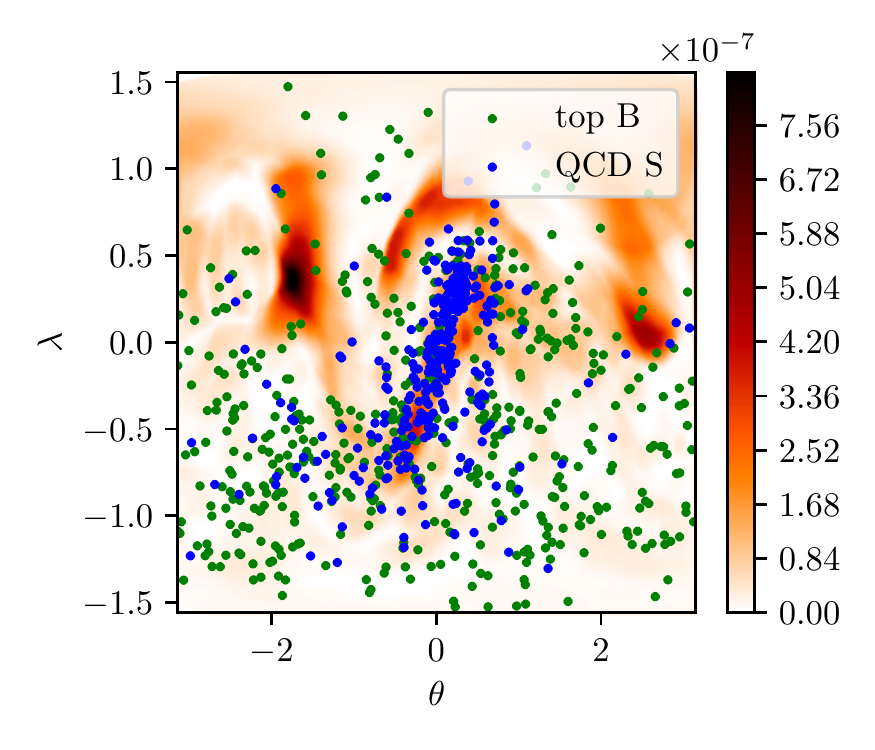} \\
  \includegraphics[width=0.495\textwidth]{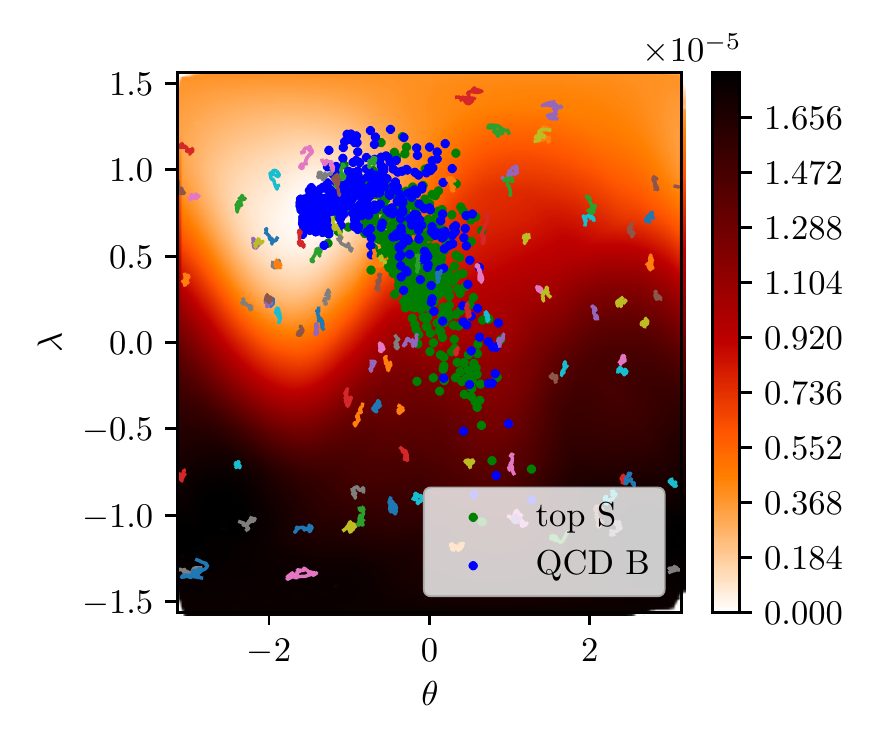}
  \includegraphics[width=0.495\textwidth]{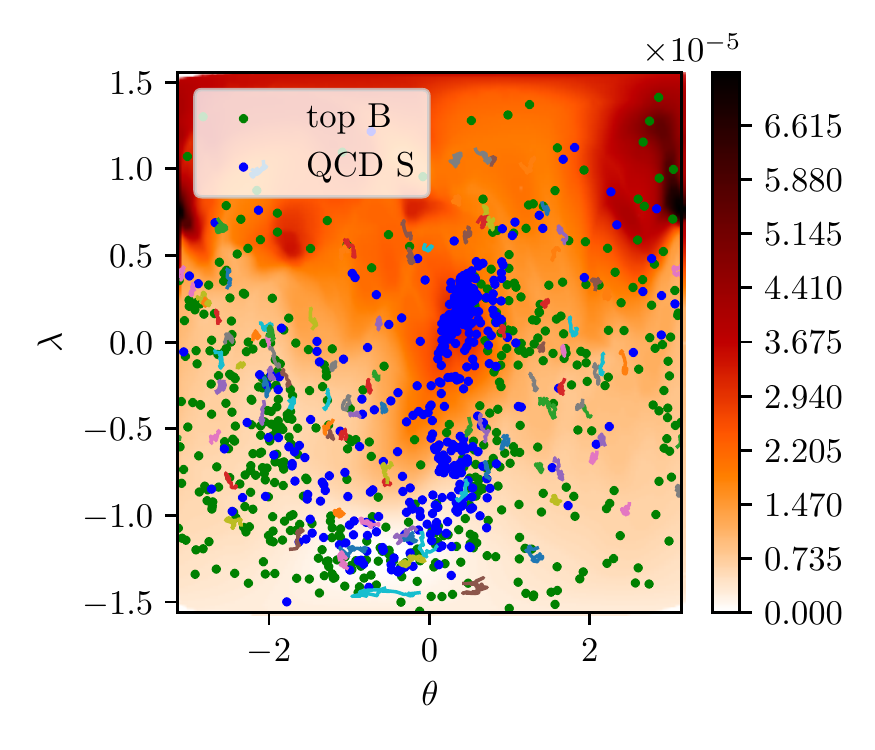}
  \caption{Equirectangular projection of the latent space after
    pre-training (upper) and after NAE training (lower). The $x$- and
    $y$-axis are the longitude and latitude on the sphere. We train on
    QCD jets (left) and on top jets (right). The background color
    indicates the energy over the latent space, the lines represent
    the path of the LMCs in the current iteration, and the points show
    the positions of jets from both samples.}
  \label{fig:top_lat}
\end{figure}
%----------------------------------------------------------

%%%MOD
A simple, established anomaly detection task based on jet images is to
extract top jets out of a QCD jet sample, with network training on
background only~\cite{Heimel:2018mkt}.  We summarize the results with
a focus on the performance for symmetric tagging of QCD vs top
images.
In standard autoencoder models a known problem is that they tend to assign 
larger reconstruction losses to samples with higher complexity rather than those 
which are not well-represented in the training data.
This is exemplified when training an autoencoder on QCD jets to identify 
anomalous top jets, versus training the autoencoder on top jets to identify 
anomalous QCD jets.
The underlying physics suggests that it should be easy to find large regions
of phase space exclusively populated by QCD or top jets, therefore allowing
for out-of-distribution detection in both directions.
However, the autoencoder works well in the direction of anomalous top jets, while it does 
not work well in the direction of anomalous QCD jets.
We refer to any anomaly detection technique that tags in both directions of higher 
and lower jet complexity, without modifications in the architecture and training, as symmetric.

A known issue in training EBMs is a potential collapse of the sampler,
detected by a diverging negative energy and a collapse of the sampled
images~\cite{du2019implicitEBM,kingma2018glow}. To find a sweet spot
between mode coverage and stability requires careful tuning of the LMC
parameters, in addition to a regularization. We only encounter this
failure when training on top jets, because the latent space undergoes
drastic changes.  To detect a collapse, we use several diagnostic
tools.  A proper training shows stable positive and negative energies,
a fluctuating loss function close to zero, and smooth variations of
the weights. In addition, we can directly look at the sampled images
saving batches after a fixed number of iterations.  The NAE training
is carried over for 50 epochs or until a collapse of the sampler
happens. Then, the best model is chosen by taking the iteration with
the loss function closest to zero and with stable positive and
negative energies.

We choose a three-dimensional latent space for our model, 
making it a sphere in three dimensions. We
exploit this low dimensionality to visualize the development of the
latent energy landscape.  We employ an equirectangular projection as shown in
Fig.~\ref{fig:top_lat}.  The $x$-axis and the $y$-axis give the
longitude and the latitude on the sphere. To reduce the distortion
around jets, the poles are chosen such that the center $(0,0)$ is
given by the region with most jets.  By sampling points on the sphere
and calculating the energy of the decoded jets we build the latent
landscapes.  In this landscape we show the path of latent LMCs and the
position of encoded jets from both distribution.

In the upper panels of Fig.~\ref{fig:top_lat} we show a projection of
the latent space after minimization of the MSE, like in the usual AE, 
but using a compact, spherical latent
space.  In the left panels we train on the simpler QCD background,
which means that the latent space has a simple structure. The QCD jets
are distributed widely over the low-energy region, while the anomalous
top jets cluster slightly away from the QCD jets. The situation
changes when we train on the more complex top jets, as shown in the
right panels. The latent MSE or energy-landscape reflects this complex
structure with many minima, and top jets spread over most of the
sphere. After the NAE training, only the regions populated by training
data have a low energy. The sampling procedure has shaped the decoder
manifold to correctly reconstruct only training jet images. For both
training directions, the Markov chains move from a uniform
distribution to mostly cover the region with low energy, leading to an
improved separation of the respective backgrounds and signals.\medskip

%----------------------------------------------------------
\begin{figure}[t]
  \includegraphics[width=0.495\textwidth]{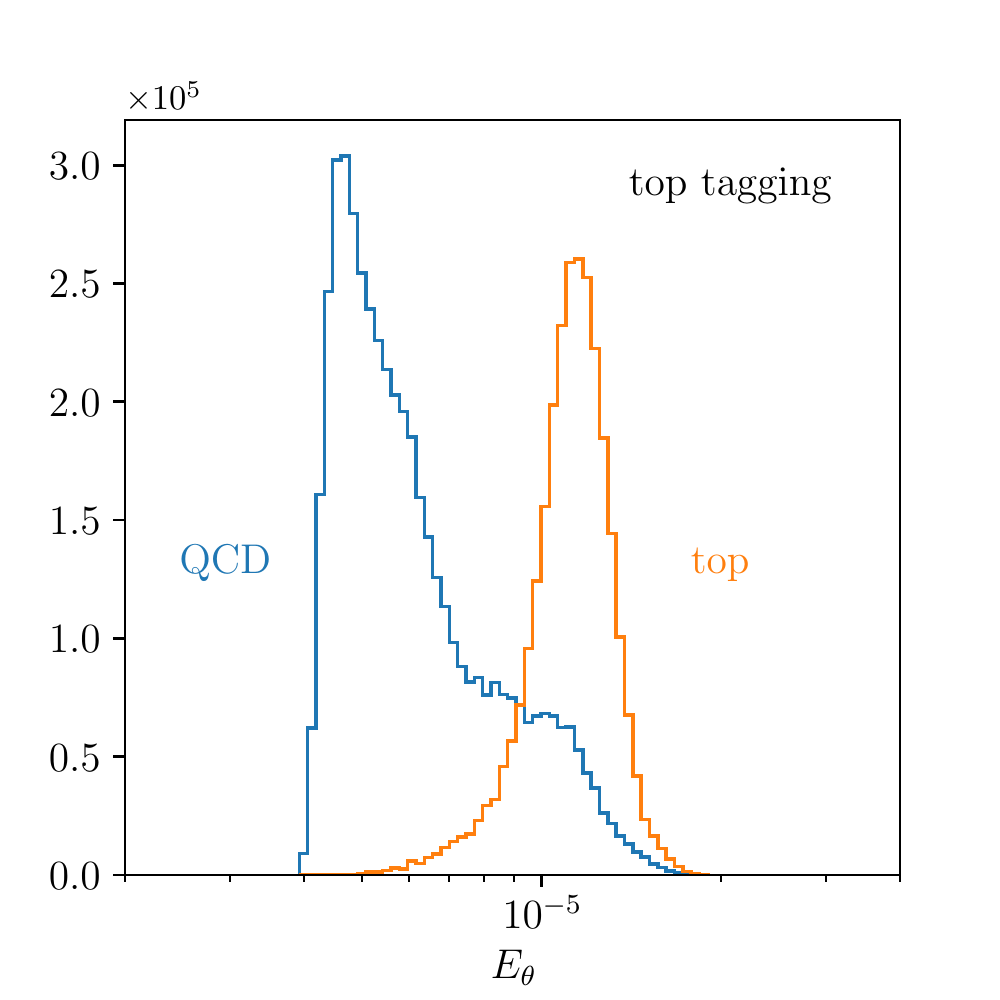}
  \includegraphics[width=0.495\textwidth]{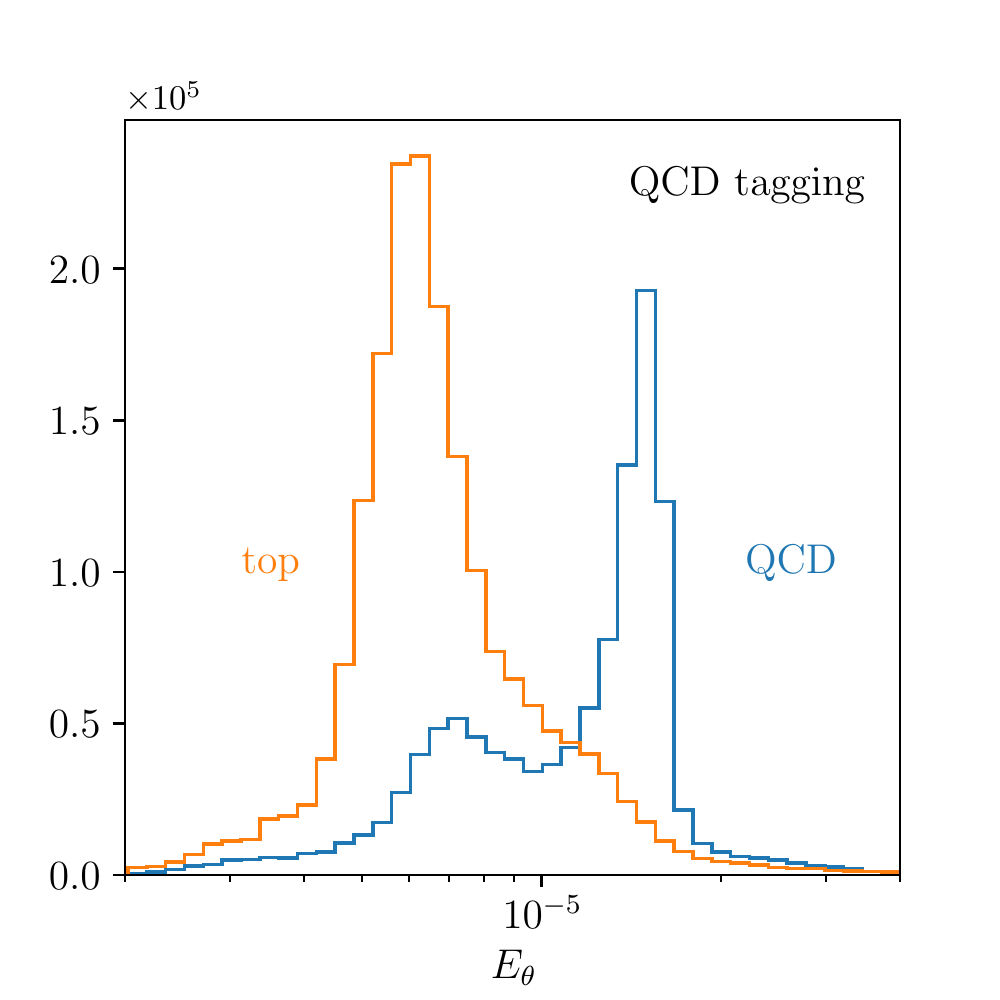}
  \caption{Distribution of the energy or MSE after training on QCD
    jets (left) and on top jets (right). We show the energy for QCD
    jets (blue) and top jets (orange) in both cases.}
  \label{fig:top_hist}
\end{figure}
%----------------------------------------------------------

%----------------------------------------------------------
\begin{figure}[t]
  \centering
  \begin{subfigure}[c]{0.495\textwidth}
    \includegraphics[width=\textwidth]{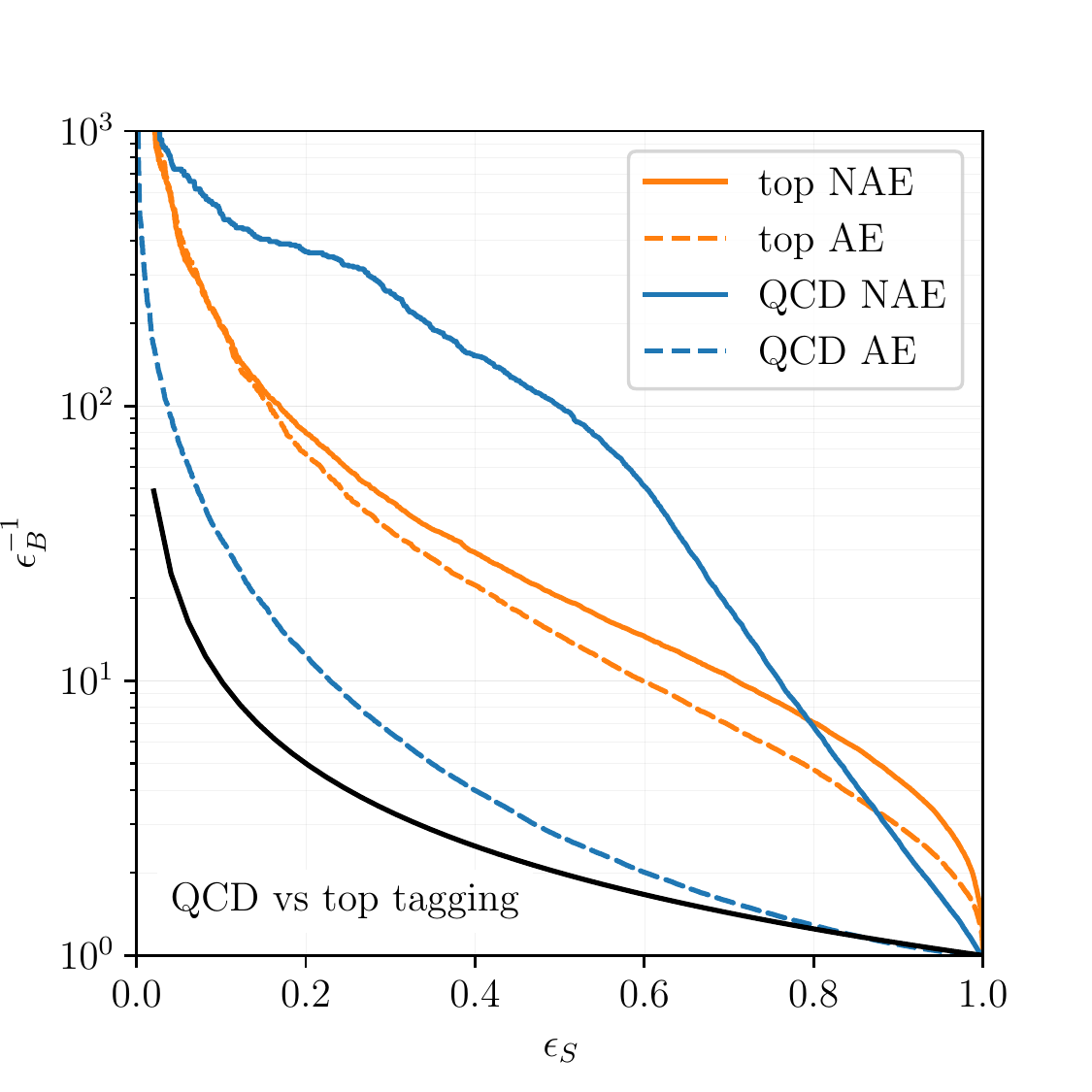}
  \end{subfigure}
  \begin{footnotesize} \begin{tabular}[c]{l|cc|c|c}
      \toprule
Signal & \multicolumn{2}{c|}{NAE} & AE~\cite{Heimel:2018mkt} & DVAE~\cite{Dillon:2021nxw} \\
 & AUC & $\epsilon_B^{-1} (\epsilon_S=0.2)$ & AUC & AUC \\\midrule
top (AE)   & 0.875 & 68  & 0.89 & 0.87 \\
top (NAE)  & 0.91 & 80 & & \\
QCD (AE)   & 0.579 & 12  & -- & 0.75 \\
QCD (NAE)  & 0.89 & 350 & &\\
\bottomrule
  \end{tabular} \end{footnotesize}
  \caption{ROC curve for top (orange) and QCD (blue) tagging after
    AE pre-training (dashed), and after NAE training (solid). A random classifier
    corresponds to the solid black line. In the
    table we compare the performance of the NAE, and the pre-trained AE
    used here, to two studies in the literature.}
  \label{fig:top_roc}
\end{figure}
%----------------------------------------------------------

To see the difference in the two-directional training we can also look
at the respective energy distributions.  In the left panel of
Fig.~\ref{fig:top_hist} we first see the result after training the NAE
on QCD jets. The energy values for the background are peaked strongly,
cut off below $4 \times 10^{-5}$ and with a smooth tail towards larger
energy values. The energy distribution for top jets is peaked at larger
values, and again with an unstructured tail into the QCD region. We
can then evaluate the performance of anomalous top tagging in terms of
the ROC curve, the AUC score, and the inverse mistag at low efficiency
($\epsilon_s=0.2$) in Fig.~\ref{fig:top_roc}. This choice of working
point is motivated by possible applications of autoencoders requiring
significant background rejection.  The orange ROC curves show how the
performance increases after the additional AE training to the NAE
training in the self-constructed latent-space geometry. The AUC value
of 0.91 quoted in the corresponding table is above the
AE setup and our earlier studies.

Next, we can see what happens when we train on top jets and search for
the simpler QCD jets as an anomaly. In the right panel of
Fig.~\ref{fig:top_hist} the background energy is much broader, with a
significant tail also towards small energy values. The QCD distribution
develops two distinct peaks, an expected peak in the tail of the top
distribution and an additional peak under the top peak. The fact that
the NAE manages to push the QCD jets towards larger energy values
indicates that the NAE works beyond the compressibility ordering of
the simple AE. However, the second peak shows that a fraction of QCD
jets look just like top jets to the NAE. The ROC curves in
Fig.~\ref{fig:top_roc} first confirm that training a regular AE to
search for QCD jets in a top sample makes little sense, leading to an
AUC value of 0.579. After the additional NAE training step we reach an
ROC value of almost 0.9, close to the corresponding value for top
tagging. However, the shape of the ROC curve does not exactly follow
our expectations. We can start with large $\epsilon_S \to 1$ in the
right panel of Fig.~\ref{fig:top_hist}. Here the working point is in
the small-energy tails of the signal and background distributions, and
because of the tails in the top jet distribution the performance of
the classification network starts poorly. Moving towards smaller
$\epsilon_S$ the network performance drastically improves, until we
pass the background peak, corresponding to $\epsilon_S \sim
0.6$. Below this value, the QCD tagging improves, again, but more
slowly than the corresponding top tagging.

Altogether, we see in the right panels of Fig.~\ref{fig:top_roc} that
the NAE combines competitive performance with symmetric tagging top
and QCD tagging. In the easier direction of top tagging it beats the
AE and DVAE benchmarks in spite of the non-optimized setup, and in the
reverse direction of QCD tagging it provides competitive results for
the first time.

%%%%%%%%%%%%%%%%%%%%%%%%%%%%%%%%%%%%%%%%%%%%%%%%%%
\section{QCD vs dark jets}
\label{sec:dark}

After testing NAE on this benchmark process, we can move to a more
difficult task, namely tagging two distinct kinds of dark jets with
the same network.  The signal datasets are the same as in
Ref.~\cite{Buss:2022lxw}.

\begin{figure}[t]
 %0.5
  \includegraphics[width=0.245\textwidth]{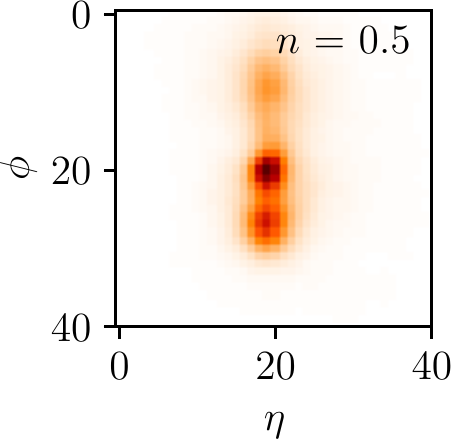}
  \includegraphics[width=0.245\textwidth]{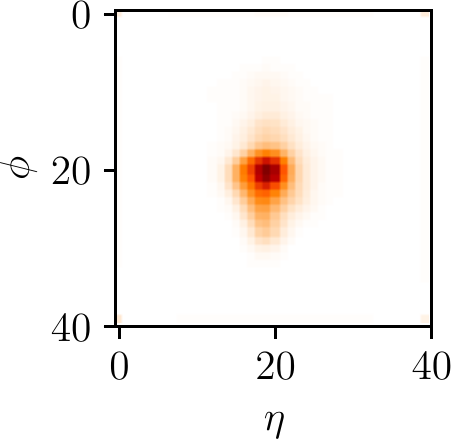}
  \includegraphics[width=0.245\textwidth]{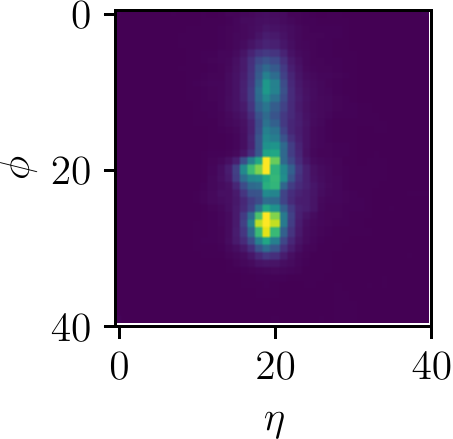}
  \includegraphics[width=0.245\textwidth]{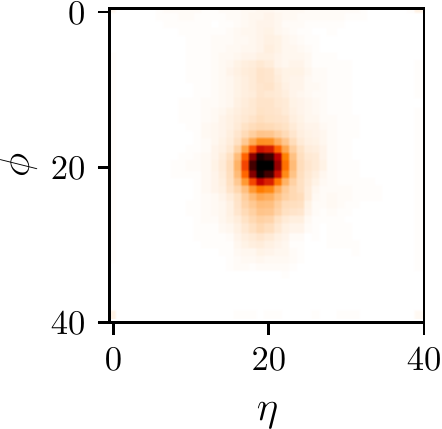} \\
 %0.2
  \includegraphics[width=0.245\textwidth]{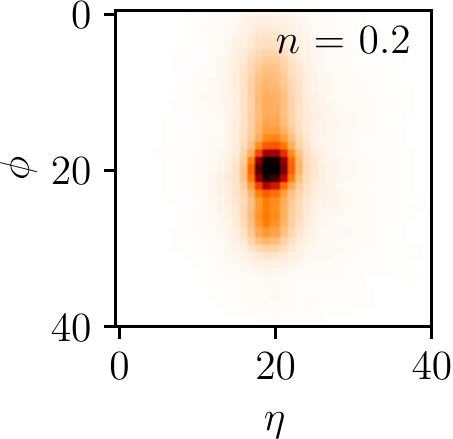}
  \includegraphics[width=0.245\textwidth]{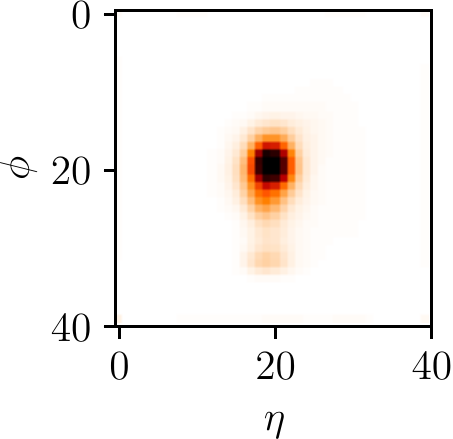}
  \includegraphics[width=0.245\textwidth]{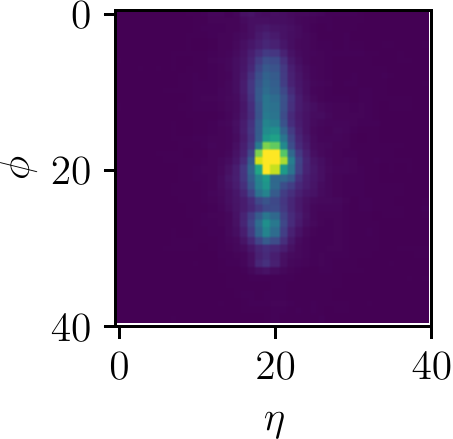}
  \includegraphics[width=0.245\textwidth]{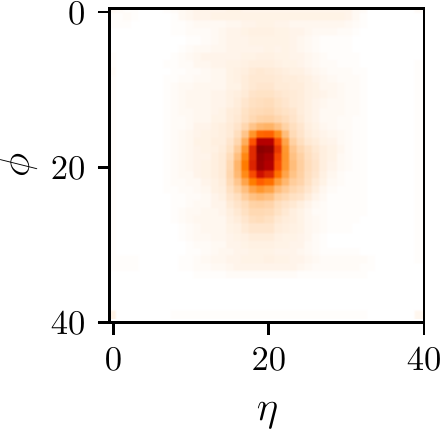} \\
 %0.01
  \includegraphics[width=0.245\textwidth]{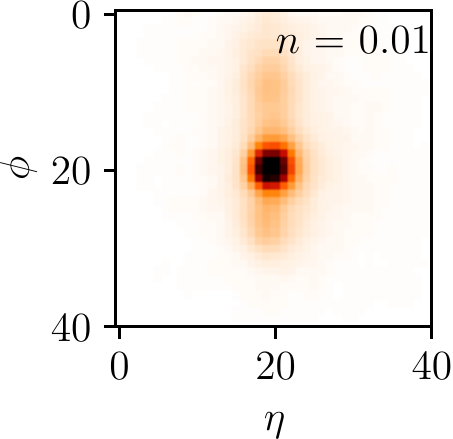}
  \includegraphics[width=0.245\textwidth]{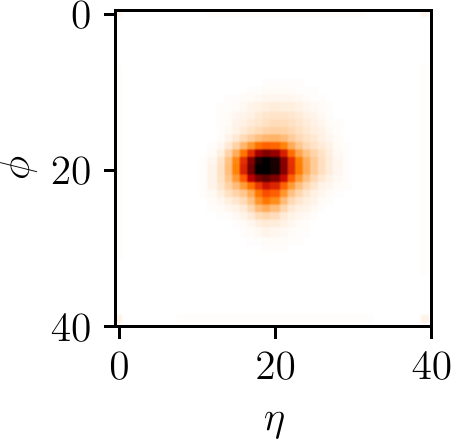}
  \includegraphics[width=0.245\textwidth]{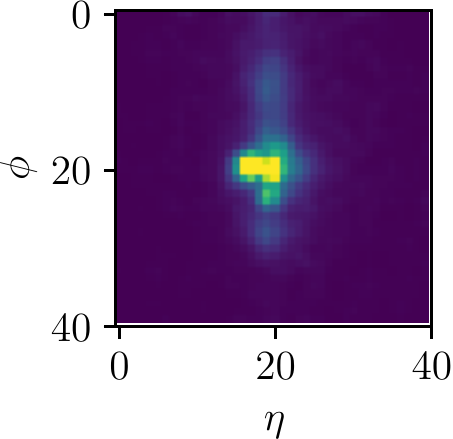}
  \includegraphics[width=0.245\textwidth]{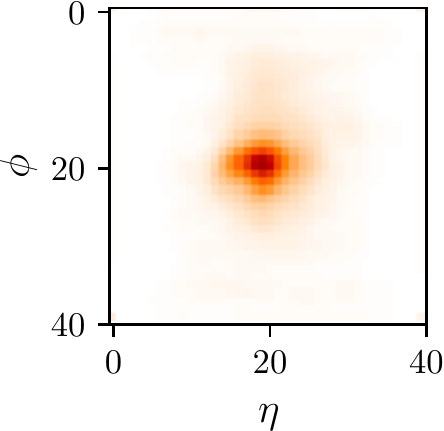} 
  \caption{Average QCD jet images for the 1k most poorly reconstructed
    jets, from left to right: average input, average reconstruction,
    pixel-wise energy between the two, and average output of the negative
    energy sample used during training in the last iteration. The rows
    correspond to reweighting factors $n=0.5, 0.2, 0.001$. }
  \label{fig:avgimg}
\end{figure}

To first illustrate the $p_T$-reweighting we select the most poorly
reconstructed 1000 QCD images, according to their MSE or energy.  In
Fig.~\ref{fig:avgimg} we show the average of these images to the left,
the average reconstruction in the second column, and the pixel-wise
energy between the two in the third row. Reducing the remapping defined
in Eq.\eqref{eq:remap} from $n = 0.5$ to $n= 0.01$ washes out the
$p_T$-structures, so the input and especially the reconstructed images
change from more structured jets to a simple, single-prong
structure. For our two signal hypotheses this means that for large $n$
the poorly reconstructed QCD images resemble the Heidelberg signal,
leading to a more efficient signal extraction, while for small $n$ the
poorly reconstructed jet images resemble the Aachen dataset.

%----------------------------------------------------------
%----------------------------------------------------------

This difference in the jet reconstruction for different models can be
explained by looking at the sampled distributions during training.
The NAE-sampled average of the negative-energy jets in the last
iteration is shown in the two right column of
Fig.~\ref{fig:avgimg}. At $n=0.5$ the NAE sampling discards all
secondary clusters and focuses on the main feature of the QCD jets,
the single prong. During training, the loss function enhances the main
feature by increasing the energy of everything around it in the latent
and phase spaces. As a consequence, the initial background is lost
after some epochs, but to keep the normalization of each jet the
central prong is enhanced.  As a result, the tagging of two-prongs
structure like the Heidelberg jets is improved.  Conversely, at
$n=0.01$ the reweighting enhances the secondary cluster, which cannot
be discarded by the training anymore. As a result, in both initial
iteration, and at equilibrium there is a residual background away from
the central prong. This way, the NAE training increases the energy for
Aachen jets, because the normalization forces the main prong to a
lower value. These effects are inevitable when training likelihood-based models,
a different preprocessing will change the density and, therefore, the anomaly score.\medskip

%----------------------------------------------------------
\begin{figure}[t]
  \includegraphics[width=0.495\textwidth]{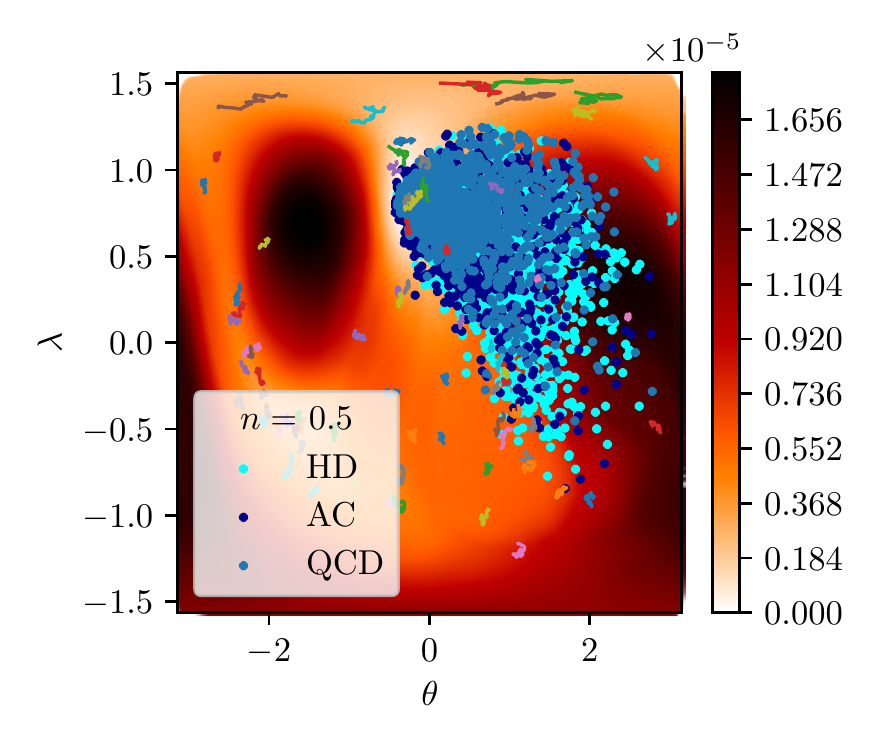}
  \includegraphics[width=0.495\textwidth]{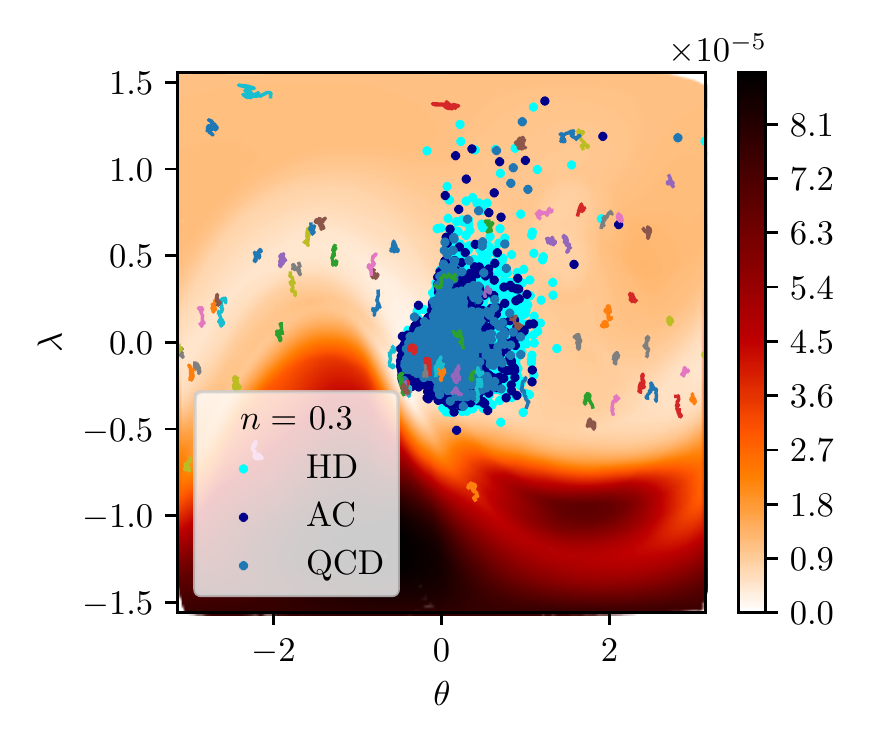}\\
  \includegraphics[width=0.495\textwidth]{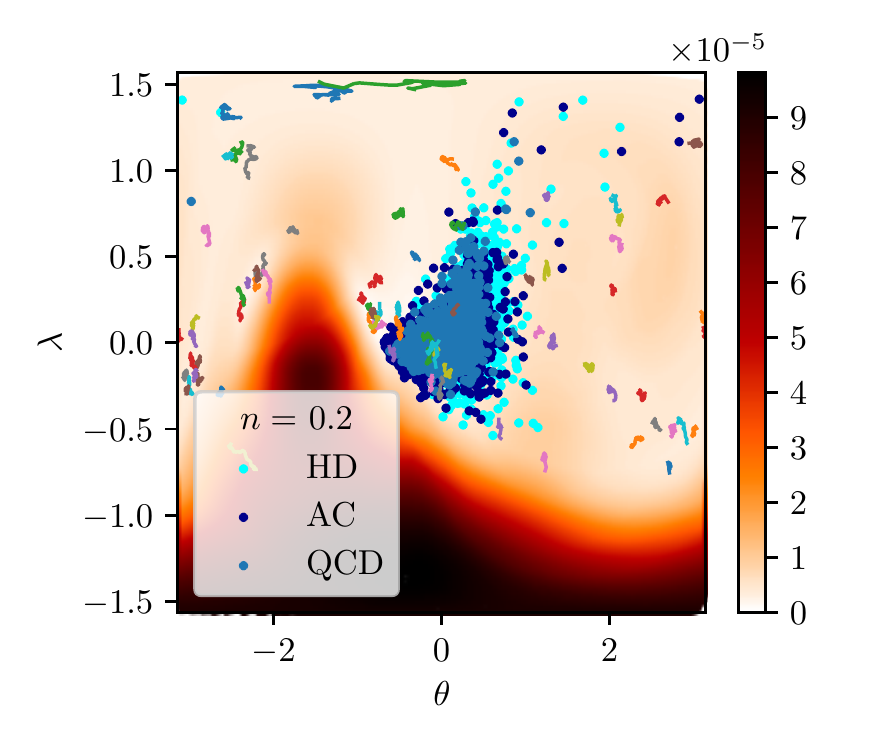}
  \includegraphics[width=0.495\textwidth]{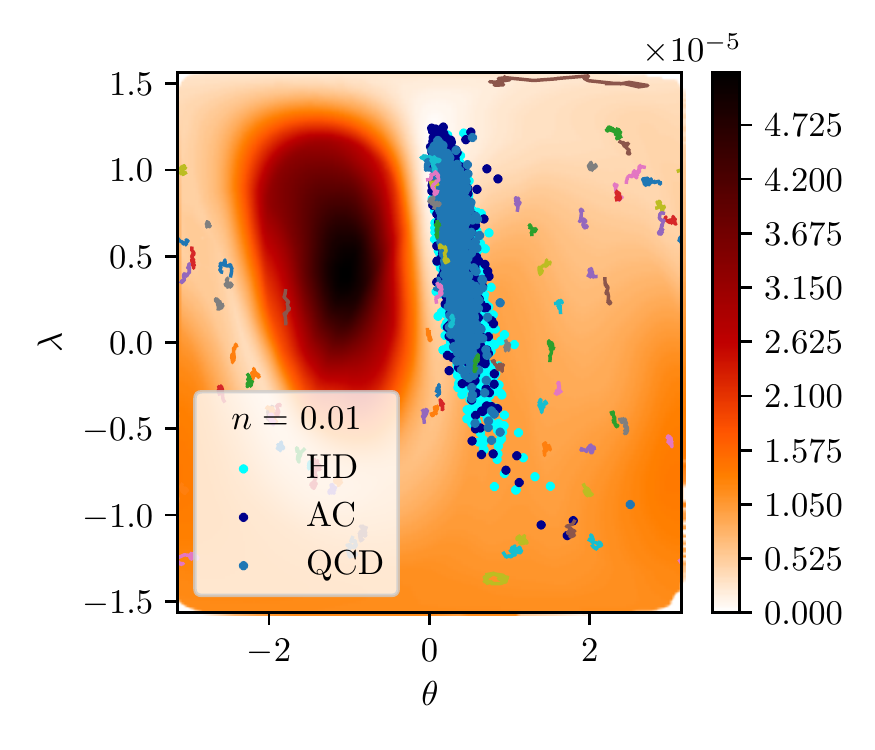}
  \caption{Equirectangular projection of latent spaces after NAE
    training on QCD jets to identify anomalous dark jets. We show
    four different $p_T$-reweightings $n$. The blue points represent a
    sub-sample of QCD training events.}
  \label{fig:djlat}
\end{figure}
%----------------------------------------------------------

As for the top vs QCD tagging, we then show the latent space
landscapes after NAE training in Fig.~\ref{fig:djlat}. For all
$p_T$-reweightings the network identifies the least populated regions
in the decoder manifold and increases the corresponding energy. As
discussed above, a large $n=0.5$ enhances the sensitivity to the more
complex Heidelberg dataset, while the sparse Aachen dataset is hardly
separated from the QCD jets. For small $n = 0.01$ a distinct region
appears at large $\lambda$, where the Aachen signal extends beyond the
QCD region.  In between the two extremes, the latent landscape changes
smoothly with the biggest change happening around $n=0.2$. Around this
point the training can oscillate between focusing on primary prongs or
on secondary clusters, causing fluctuations in the performance. This
transition in the $p_T$-reweighting is also the only case where the
hyperparameters, and especially the temperatures, have a noteworthy
effect on the network performance.

%----------------------------------------------------------
\begin{figure}[t]
  \begin{subfigure}[c]{0.32\textwidth}
    \includegraphics[width=\textwidth]{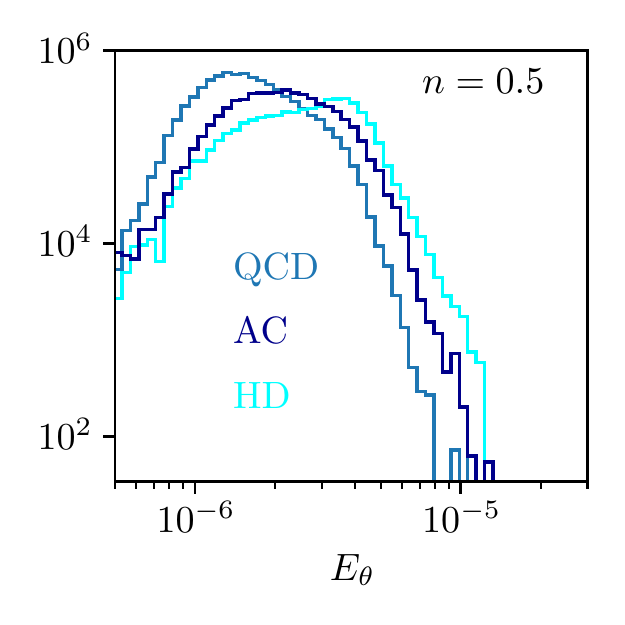}
  \end{subfigure}
  \begin{subfigure}[c]{0.32\textwidth}
    \includegraphics[width=\textwidth]{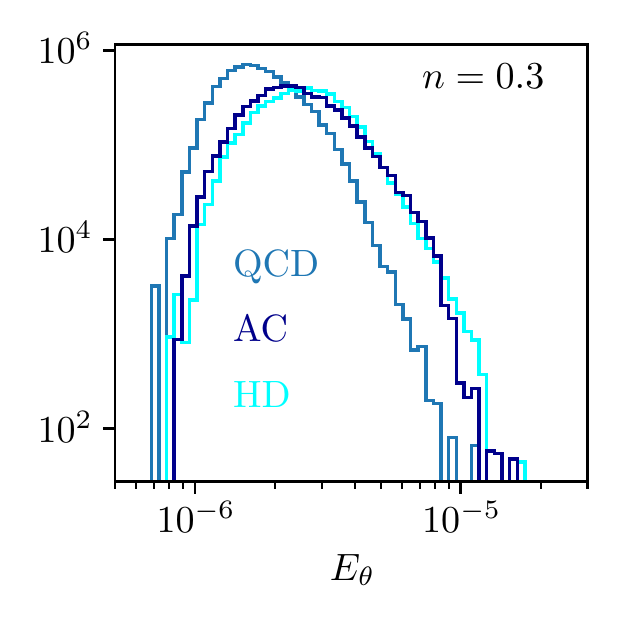}
  \end{subfigure}
  \begin{subfigure}[c]{0.32\textwidth}
    \includegraphics[width=\textwidth]{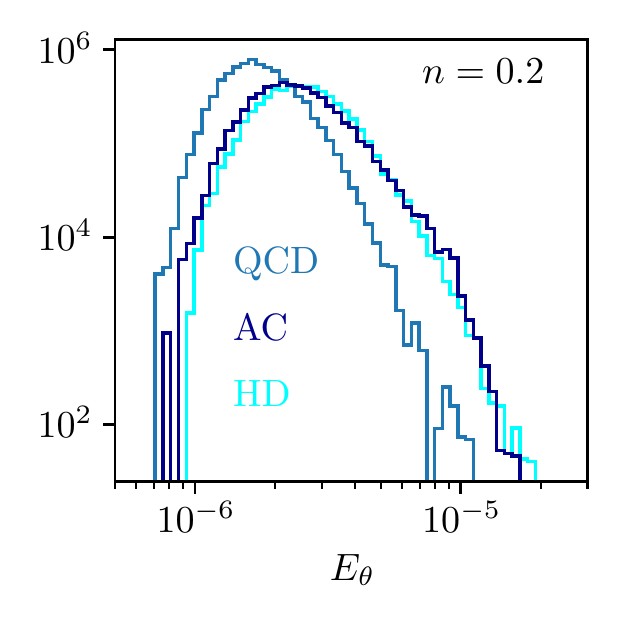}
  \end{subfigure}
  \begin{subfigure}[c]{0.32\textwidth}
    \includegraphics[width=\textwidth]{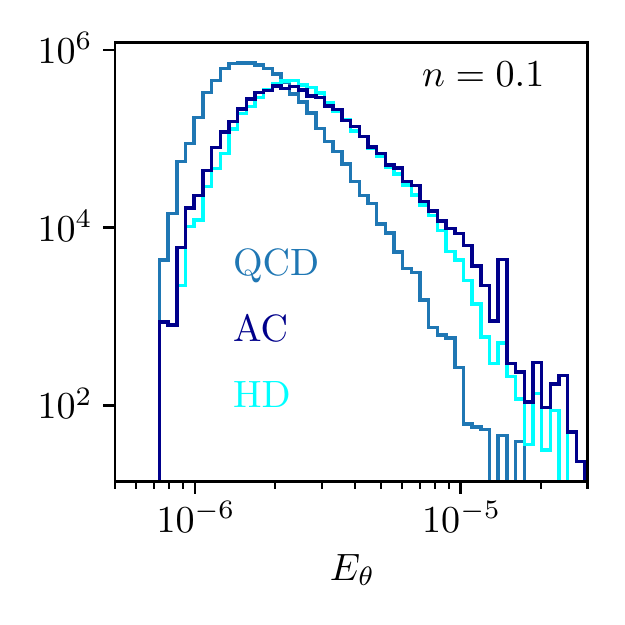}
  \end{subfigure}
  \begin{subfigure}[c]{0.32\textwidth}
    \includegraphics[width=\textwidth]{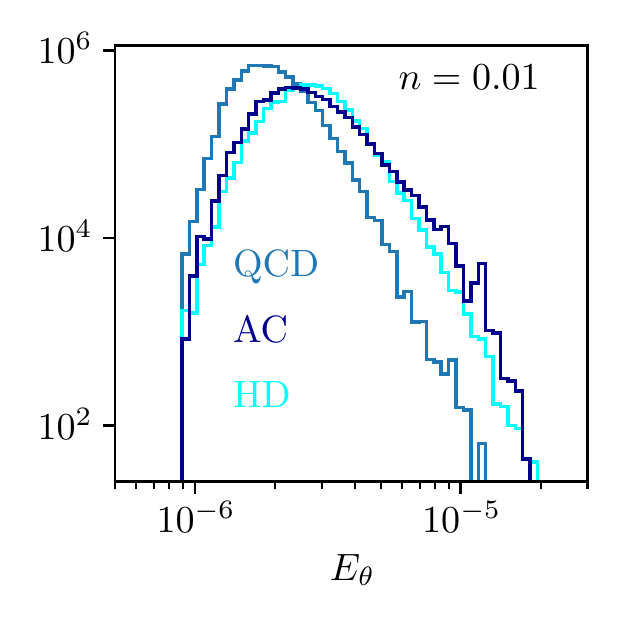}
  \end{subfigure}
  \resizebox{0.32\textwidth}{!}{%
    \begin{small}\begin{tabular}[c]{c|c|c|c|c|c|c}\toprule
  & \multicolumn{2}{c}{QCD} & \multicolumn{2}{c}{HD} & \multicolumn{2}{c}{AC}  \\
 $n$ & $\mu $ & $\sigma$ & $\mu $ &$\sigma$ & $\mu $ & $\sigma$ \\\midrule
0.5 & 2.0 &   0.9    & 3.3 &   1.4    & 2.7 &  1.1 \\
0.3 & 2.0 &   0.8   & 3.1 &    1.3   & 3.0 &   1.3 \\
0.2 & 2.0 &   0.8    & 3.1 &   1.3   & 3.0 &   1.5 \\
0.1 & 2.0 &    0.9   & 3.2 &   1.7  & 3.4 &   2.2 \\
0.01 & 2.2 &  0.8   & 3.3 &   1.4  & 3.4 &  1.8 \\\bottomrule
    \end{tabular} \end{small}
	}
  \caption{Distribution of the energy for QCD, Aachen, and Heidelberg
    datasets. Each panel corresponds to a different reweighting of the
    same datasets.  The table shows the mean and the standard
    deviation for each distribution ($\times 10^{-6}$).}
  \label{fig:djmse}
\end{figure}
%----------------------------------------------------------

%----------------------------------------------------------
\begin{figure}[b!]
  \includegraphics[width=0.48\textwidth]{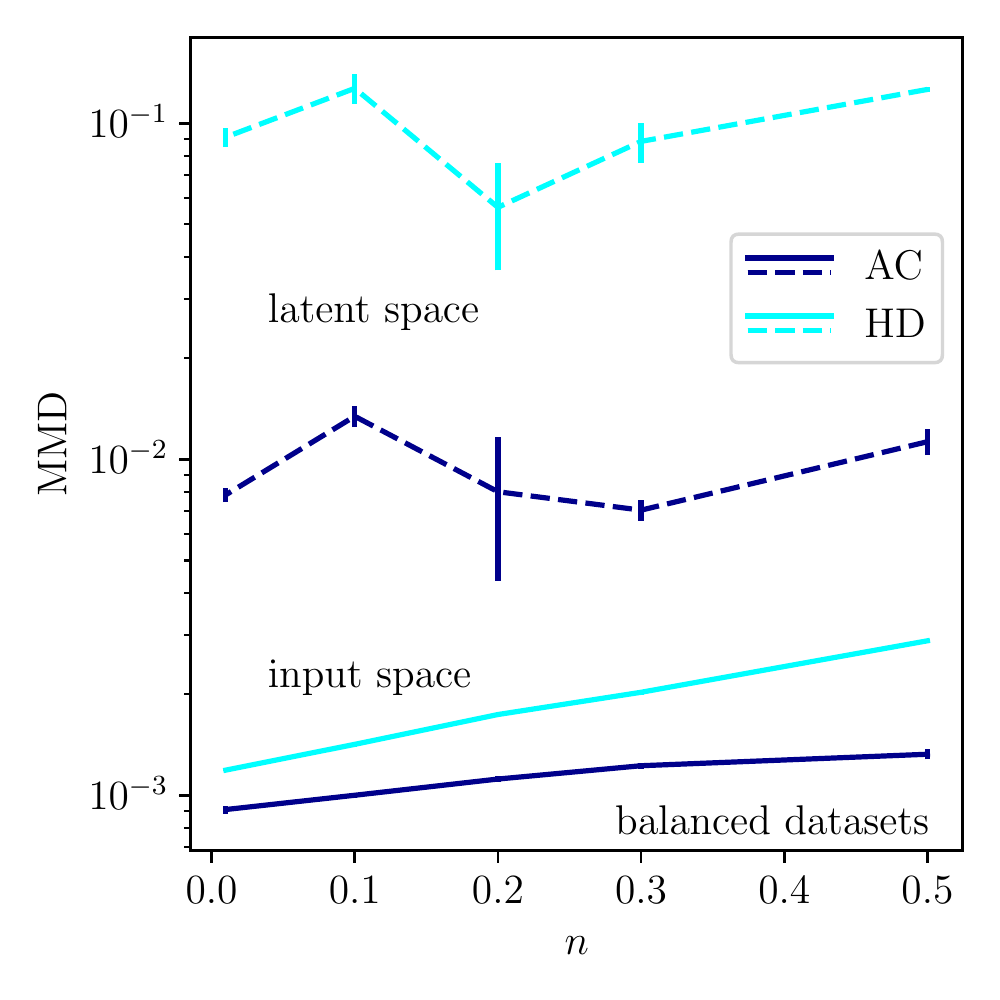}
  \includegraphics[width=0.48\textwidth]{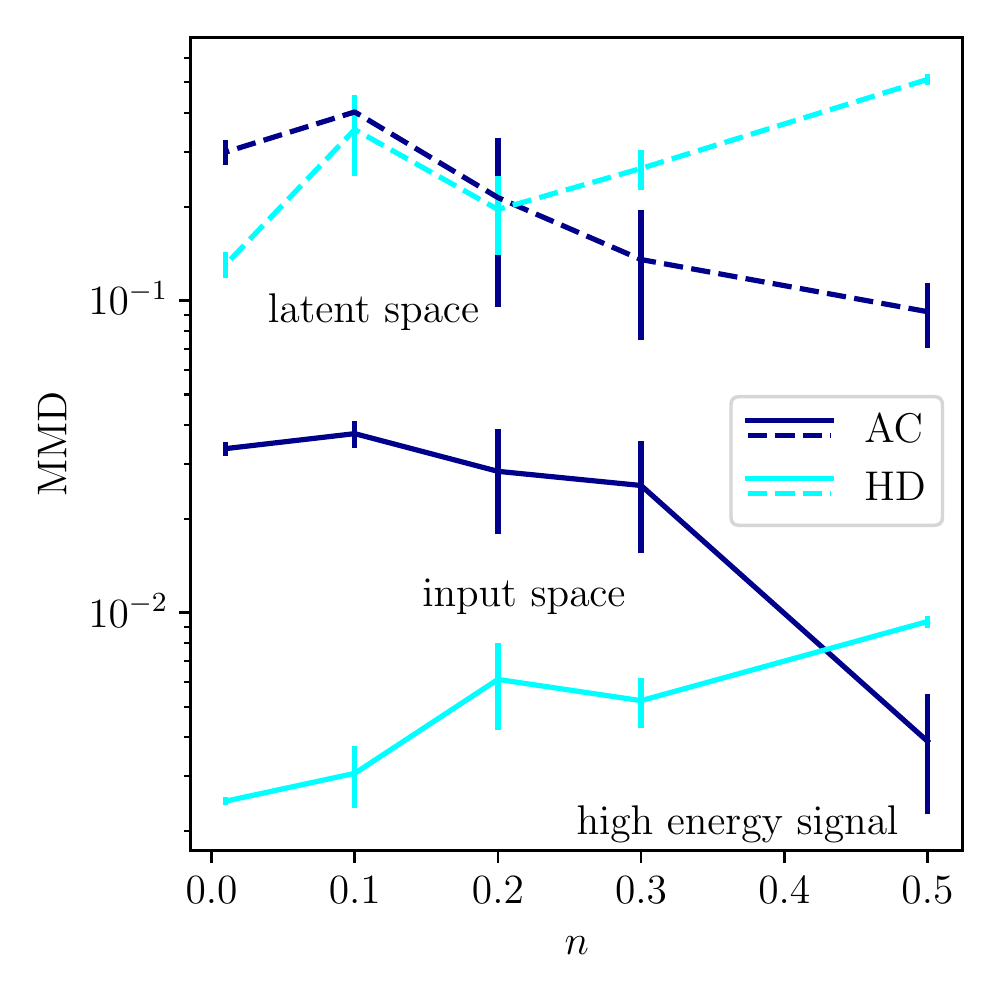}
  \caption{MMD in phase (solid) and latent (dashed) space between two
    random sample of QCD and signal jets (left), and between the QCD
    sample and the most poorly reconstructed signal jets
    (right). Unlike for the other figures, the remapping $n$ increases
    from left to right.}
  \label{fig:mmdfull}
\end{figure}
%----------------------------------------------------------

Once again focusing on the $p_T$-reweightings we show the energy
distributions for the QCD training data and the two signals in Fig.~\ref{fig:djmse}.
We see that unlike in our earlier study~\cite{Buss:2022lxw} the effect of the preprocessing
on the whole distribution is limited. A shift in performance at low signal
efficiency can be seen by varying $n$ with the ordering between 
the two dataset being switched around $n = 0.2~...~0.3$.
The energy distribution of the Heidelberg dataset has a shifted
main peak at $n=0.5$ which is washed out by smaller reweighting factors,
while the QCD distribution undergoes a slight shift and develops a
longer high-energy tail. For the Aachen dataset, lowering $n$ moves the mean away 
from the QCD background and at the same time increases the width of the distribution.
These patterns will affect the ROC curves at low signal efficiency and large background
suppression.

Once we understand how the $p_T$-reweighting changes the input
distributions and the energy distributions for the QCD background and
the dark jets signals, we can measure the difference between QCD jets
and each of the two signals by computing the maximum mean discrepancy
(MMD)~\cite{gretton2012kernel} of sub-samples from these
distributions.  We show these MMD curves as a function of $n$ in
Fig.~\ref{fig:mmdfull}, for the full distributions of 20k QCD and
signal jets and only considering the 1k most poorly reconstructed jets
in the high-background-rejection target region. For the input
distributions we see the same trend in both panels --- small $n$
benefits the tagging performance for the Aachen dataset and decreases
for the Heidelberg dataset; increasing $n$ improves the tagging
performance for the Heidelberg dataset.

The more interesting question is if the input-space pattern can also
be observed in the latent-space distributions. The idea behind this
test is to construct an autoencoder with a choice of anomaly scores,
either reconstruction-based in phase space or in the latent
space~\cite{Dillon:2021nxw}. Again in Fig.~\ref{fig:mmdfull} we show
the corresponding MMD values as dashed curves. For the complete
samples as well as for the most poorly reconstructed jets the
latent-space MMD behaves just like the phase-space MMD.  This
indicates that it should, if necessary, be possible to construct a
latent-space anomaly score for the NAE.\medskip

%----------------------------------------------------------
\begin{figure}[t]
  \includegraphics[width=0.495\textwidth]{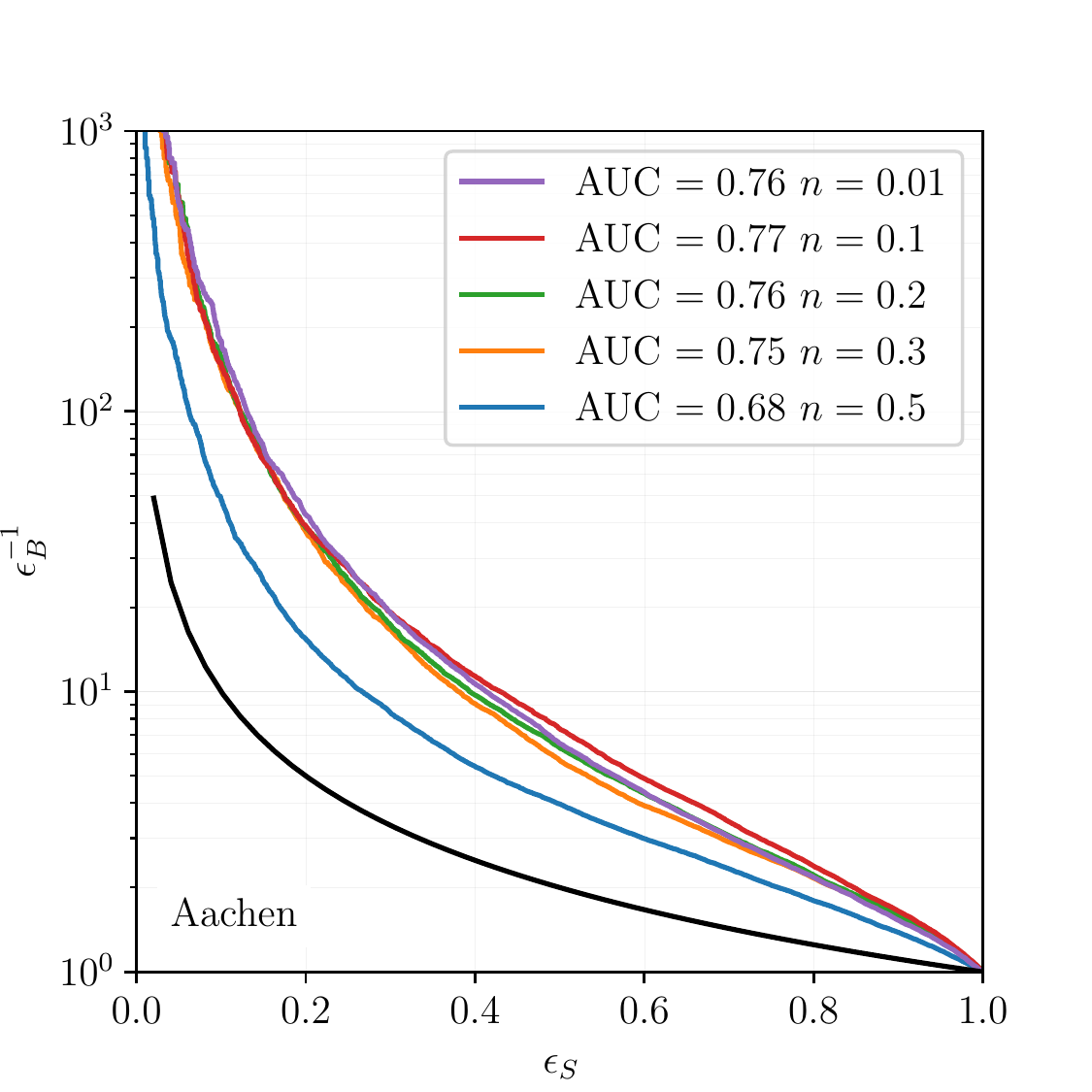}
  \includegraphics[width=0.495\textwidth]{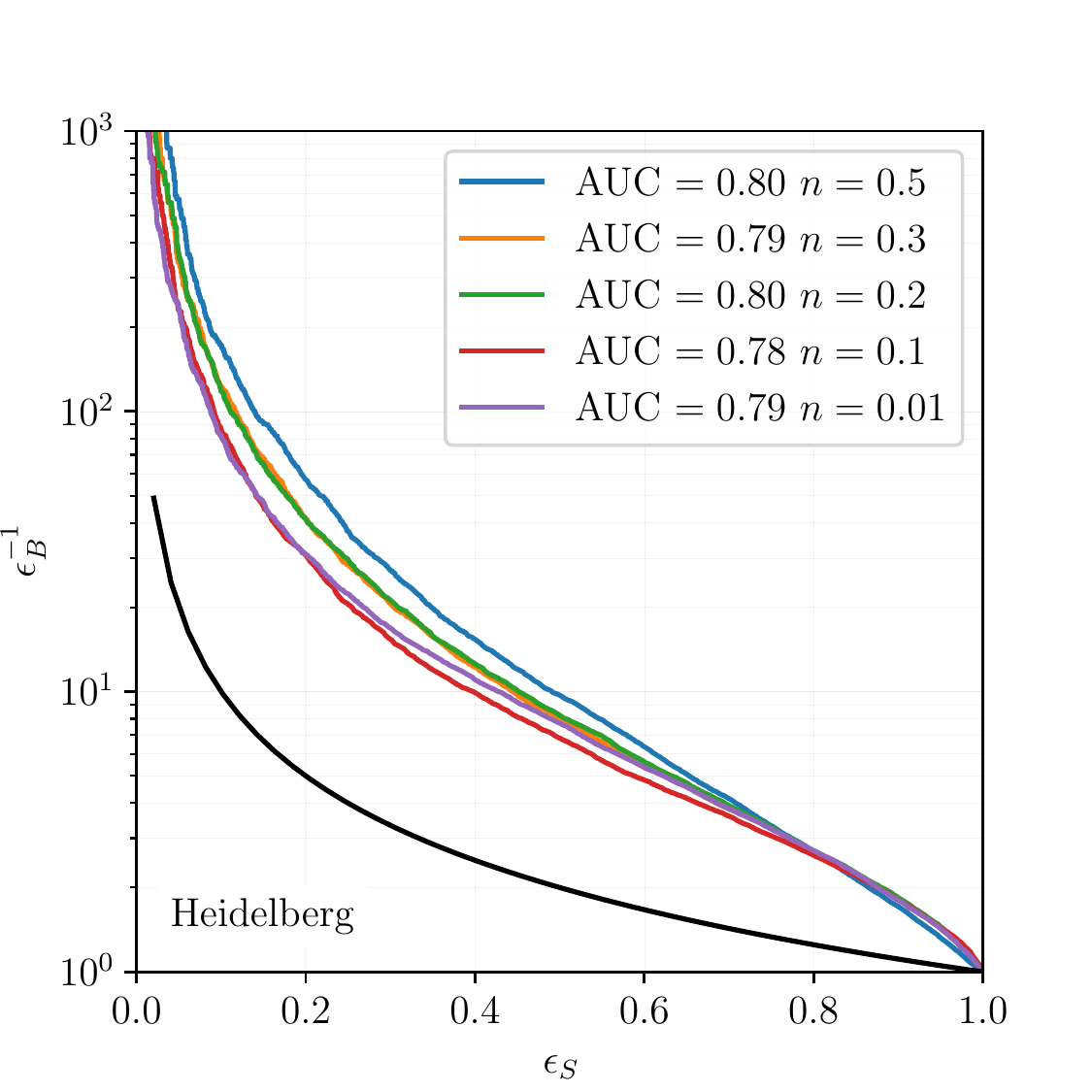} \\[3ex]
  \centering
  \small{
  \begin{tabular}[c]{l@{\hskip 0.1in}@{\hskip 0.1in}c@{\hskip 0.1in}|@{\hskip 0.1in}c@{\hskip 0.1in}|@{\hskip 0.1in}c
  				@{\hskip 0.1in}|@{\hskip 0.1in}c@{\hskip 0.1in}|
				@{\hskip 0.1in}c@{\hskip 0.1in}|@{\hskip 0.1in}c@{\hskip 0.1in}} \toprule
  & & \multicolumn{5}{c}{$n$}  \\
  Data & & 0.5 & 0.3 & 0.2 & 0.1 & 0.01 \\ \midrule
  \multirow{2}{*}{Heidelberg} & AUC &        0.795 (5) & 0.796 (5) & 0.789 (8) & 0.78 (1)  & 0.790 (5) \\ \rule{0pt}{2ex}     
  &  $\epsilon_B^{-1} (\epsilon_S=0.2)$ &  62 (3)       & 42 (5)      &  42 (4)    & 28 (4)     & 30 (1) \\ \midrule
  \multirow{2}{*}{Aachen}  & AUC &           0.68 (1)    & 0.746 (5) &  0.75 (1)  & 0.767 (5) & 0.755 (5) \\ \rule{0pt}{2ex}   
  &  $\epsilon_B^{-1} (\epsilon_S=0.2)$ & 15 (1)       & 38 (3)      &      33 (7) & 41 (2)      & 41 (1) \\ \bottomrule 
  
  \end{tabular} }
  \caption{ROC curve for dark jets tagging with different reweightings
    $n$, shown for the Aachen signal (left) and the Heidelberg signal
    (right). A random classifier corresponds to the solid black line. 
    The table is based on the same information and shows the 
    mean and the standard deviation of five different runs.}
  \label{fig:djrocrew}
\end{figure}
%----------------------------------------------------------

Moving on to the performance of the NAE on dark jets, we show the ROC
curves with different reweightings in Fig.~\ref{fig:djrocrew}. First,
we see that the AUCs for the Aachen and Heidelberg datasets are
roughly similar.  For the sparse Aachen jets we already know that
smaller values of $n$ benefit the tagging performance, but we also see
that for $n < 0.3$ the AUC reaches values above 0.72, and for
$n=0.2~...~0.01$ the performance essentially plateaus at a high
level. In contrast, for the Heidelberg signal we expect a better
tagging performance around $\epsilon_S\ \sim 0.2$ for larger
$n$-values.

From Fig.\ref{fig:djmse} we know that the different reweightings
mostly change the ordering of the two signal tails at high energies
and leave the bulks of the distributions unchanged.  The corresponding
ROC curves in Fig.~\ref{fig:djrocrew} confirm that the remaining
$n$-dependence is connected to a behavioral change in the model in the
region $n \sim 0.2$. While the choice $n=0.2$ is not optimal for each
of the signals, it can be used as a working compromise between sparse
dark jets and dark jets related to a mass drop.

%%%%%%%%%%%%%%%%%%%%%%%%%%%%%%%%%%%%%%%%%%%%%%%%%%%%%%%%%%%%%%%%%
\section{Outlook}

%%%MOD
Autoencoders are ML-analysis tools which effectively represent the idea
behind LHC searches. Unsupervised training can conceptually enrich
many aspects of LHC physics, from trigger to analysis techniques.
Standard autoencoders identify out-of-distribution jets or events
based on the compressibility of their features, a method which is typically not applicable to LHC physics. They are also
closely tied to the compressibility of jets or events, a bias we need
to avoid because new physics can be more or less complex than QCD. In
practice this means that autoencoders should identify backgrounds and
anomalies symmetrically.  An alternative strategy to define anomalies
is based on regions with low phase space densities.
The main goal of this work is to define an autoencoder which can 
reliably identify anomalous jets with an improved training procedure.
However, density-based autoencoders by definition have a dependence 
on data preprocessing~\cite{Buss:2022lxw}.
The additional components of the NAE architecture
do not increase the size of the network or the inference time, they only increase
the time taken to train the model.

The NAE combines a standard autoencoder architecture with an
energy-based normalization in the loss. This means it constructs an
energy or MSE landscape such that any anomaly with features not
present in the training data will be pushed to even larger
energies. Because of the normalization, the NAE can also balance
different kinds of features, which means that also the absence of a
background feature in signal jets will be visible in the energy
landscape. Working on a compact latent space, the NAE can be
understood as an extrapolation of the energy distribution beyond the
regions defined by the background sample.

Technically, the NAE is just a simple, small autoencoder network.  All
we adjust is the training after an AE pre-training step.  Applied to
top vs QCD jets we first show that for this extreme case of different
compressibilities the NAE still tags complex top jets in a simple QCD
background as well as simple QCD jets in a complex QCD background. In
addition, the NAE beats a standard AE as well as the advanced DVAE in
top tagging performance.

For the more challenging Aachen and Heidelberg dark jets the NAE works
 for a reasonable single choice of preprocessing. The
performance gain from using different preprocessings on the two
datasets is  explainable in terms of the changes induced on the features 
of the two signals. We did not pursue this option further, but we
see that it should be possible to construct a latent-space anomaly
score for the NAE.  While our studies indicate that the NAE is the
best-performing autoencoder to date, our setup is optimized for tests
and visualization. This means we use a 3-dimensional latent space with
a 2-dimensional sphere to track and illustrate the NAE training
progress. The next step will be to benchmark our architecture on more 
realistic datasets, optimize it for performace, and study the 
possibility of implementing the network on hardware for 
online triggering.

%%%%%%%%%%%%%%%%%%%%%%%%%%%%%%%%%%%%%%%%%%%%%%%%%%
\section*{Acknowledgments}

First, we would like to thank Ullrich K\"othe for many inspiring
discussions on neural network architectures. We are also grateful to
the Mainz Institute for Theoretical Physics, where this paper was
finalized. LF would like to thank Sangwoong Yoon for his constant and reliable support. 
This research is supported by the Deutsche
Forschungsgemeinschaft (DFG, German Research Foundation) under grant
396021762 -- TRR~257: \textsl{Particle Physics Phenomenology after the
  Higgs Discovery} and through Germany's Excellence Strategy EXC
2181/1 - 390900948 (the Heidelberg STRUCTURES Excellence Cluster).

\clearpage
%%%%%%%%%%%%%%%%%%%%%%%%%%%%%%%%%%%%%%%%%%%%%%%%%%
\appendix
\section{Appendix}
\subsection*{Network architecture}
\label{sec:network}

\begin{table}[b]
\centering
\begin{tabular}{l@{\hskip 0.2in}|@{\hskip 0.2in}l}
\toprule
Encoder         &   \makecell[l]{Conv2d(1, 8, 3, 1, 1, True) - PReLU - \\ 
					Conv2d(8, 8, 3, 1, 1, True) - PReLU - MaxPool2d(2, 2) - \\ 
					Conv2d(8, 8, 3, 1, 1, True) - PReLU - \\
					Conv2d(8, 8, 3, 1, 1, True) - PReLU - \\
					Conv2d(8, 1, 3, 1, 1, True) - PReLU - Flatten - \\
					Dense(400, 100, True) - PReLU - Dense(100, $D_\mathbf{z}$, True)} \\ \midrule
Decoder         &   \makecell[l]{Dense($D_\mathbf{z}$, 100, True) - PReLU - Dense(100, 400, True) - PReLU - \\
					Reshape(1, 20, 20) - Deconv2d(1, 8, 3, 1, 1, True) - PReLU - \\
					Deconv2d(8, 8, 3, 1, 1, True) - PReLU - \\
					Upsampling(2, 'b') - Deconv2d(8, 8, 3, 1, 1, True) - PReLU - \\
					Deconv2d(8, 1, 3, 1, 1, True) - Sigmoid\\} \\ \bottomrule
\end{tabular}
\caption{Network architecture. The latent space is $D_\mathbf{z} = 3$.}
\label{tab:net}
\end{table}

In this section we provide the details of the network architecture 
and the parameters setting. Tab.~\ref{tab:net} summarizes the layers
used in both encoder and decoder and their parameters.
The layers are defined as:
\begin{itemize}
\item[-] Conv2d(\texttt{in\_ch}, \texttt{out\_ch}, 
	\texttt{filter\_size}, \texttt{stride}, \texttt{padding}, \texttt{bias}): 2d convolutional layer;

\item[-] Deconv2d(\texttt{in\_ch}, \texttt{out\_ch}, 
	\texttt{filter\_size}, \texttt{stride}, \texttt{padding}, \texttt{bias}): 2d transposed convolutional layer;

\item[-] MaxPool2d(\texttt{filter\_size}, \texttt{stride}): 2d max-pooling layer;
\item[-] PReLU: rectified linear unit with learnable slope;
\item[-] Dense(\texttt{in\_nodes}, \texttt{out\_nodes}, \texttt{bias}): fully connected layer;
\item[-] Upsample(\texttt{scale\_factor}, \texttt{mode}): upsampling layer with bilinear algorithm;
\item[-] Flatten: flattens the input into a single dimension;
\item[-] Sigmoid: sigmoid activation function.
\end{itemize}
 
The architecture of the AE is summarized in Tab.~\ref{tab:net}.
Each layer is regularized using Spectral Normalization. 
The output of the last encoder layer is mapped to the surface 
of a hyper-sphere $\mathbb{S}^{D_\mathbf{z}-1}$. During training,
each step of the preliminary LMC is projected on the surface.
The initial latent distribution is uniform but a buffer of size 10000
is used to store the final points of each chain. Then, the initial points
are sampled from the uniform distribution with probability 0.05 or
from the buffer with probability 0.95.

Additional regularization terms are used to improve training stability. 
The L2 norm of the weights for both encoder and decoder is added
to the loss function with a coefficient $10^{-8}$. We also prevent the 
negative energy divergence by adding the average squared 
energy of the training batch to the final loss function.

The bottleneck of the training procedure is the sampling algorithm.
The parameters of the LMCs have been tuned to give fine samples after 
a small amount of steps to train a model in less than 15 hours. 
We have found that the structure of the initial manifold after pre-training
plays an important role for the following NAE updates. A large batch size
gave the best results while a smaller one during NAE showed more stable results.

%%%%%%%%%%%%%%%%%%%%%%%%%%%%%%%%%%%%%%%%%%%%%%%%%
\subsection*{Jets reconstruction and sampling}
\label{sec:jetsreco}

\begin{figure}
	\includegraphics[width=0.5\textwidth]{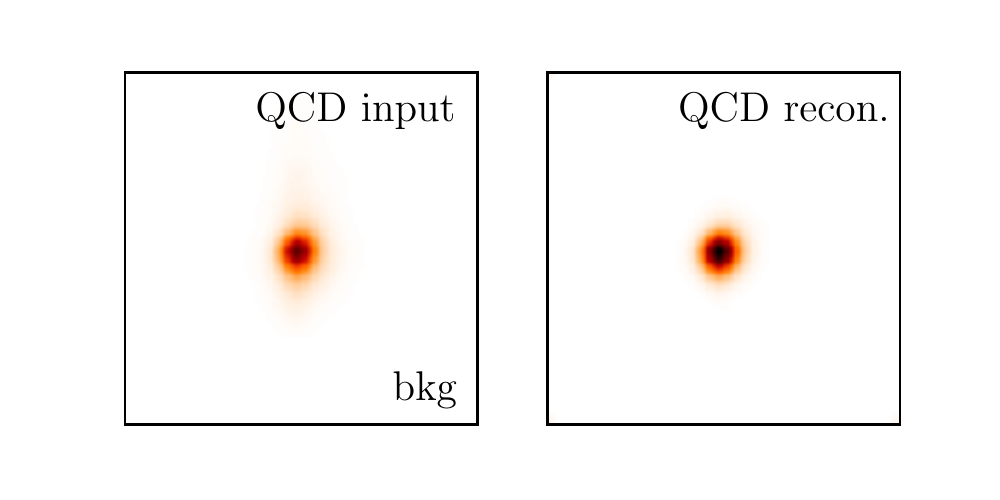} 
	\includegraphics[width=0.5\textwidth]{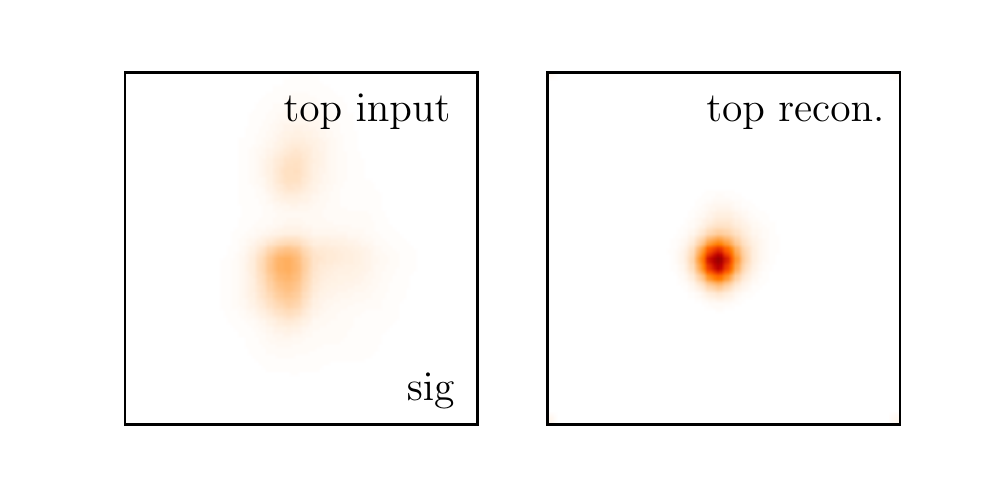} \\
	\includegraphics[width=0.5\textwidth]{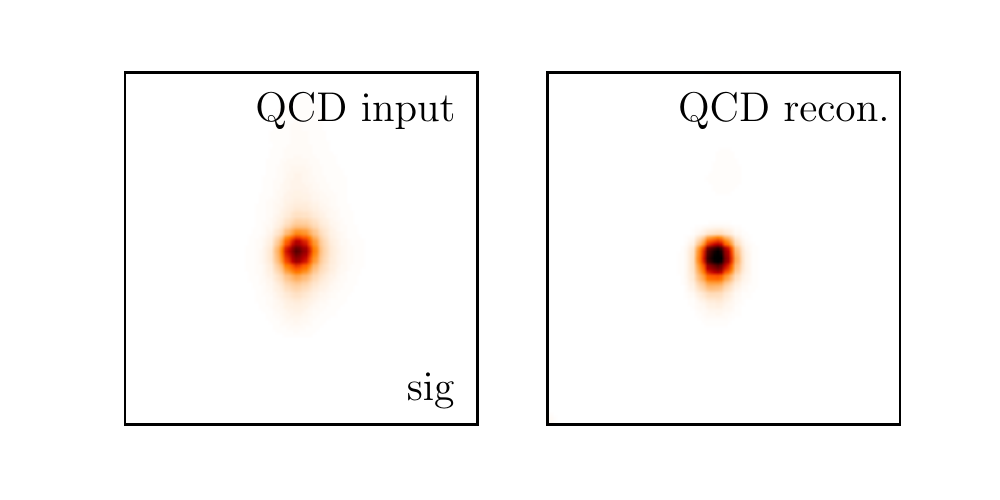} 
	\includegraphics[width=0.5\textwidth]{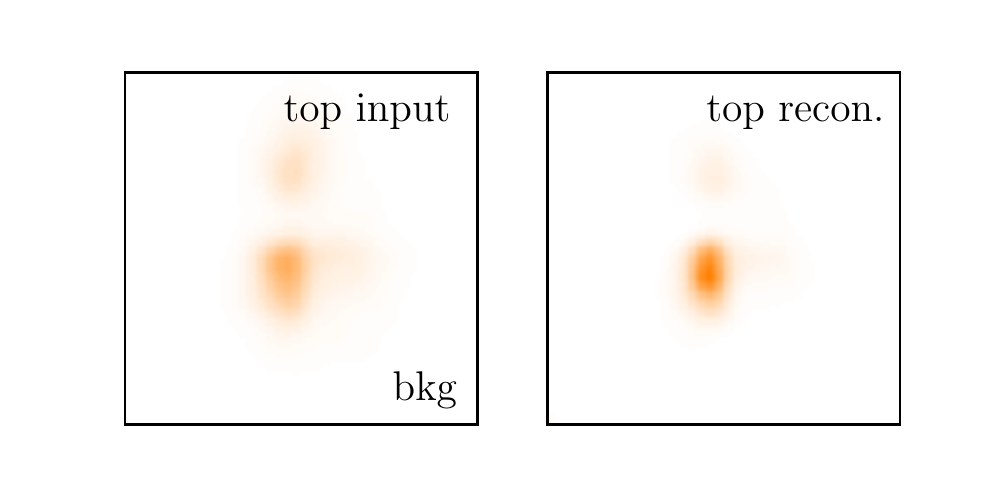} \\
	\begin{subfigure}[t]{0.5\textwidth}
	\centering
	\includegraphics[width=0.5\textwidth]{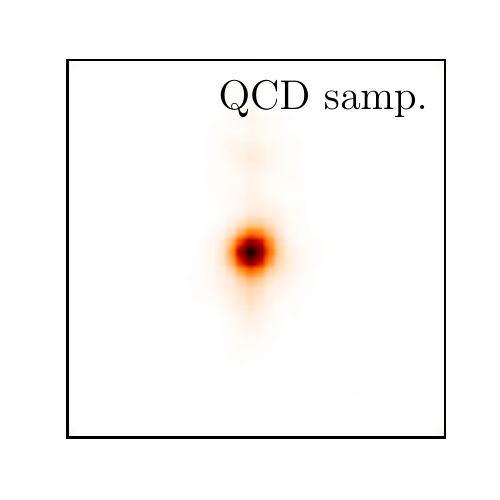}
	\end{subfigure}
	\begin{subfigure}[t]{0.5\textwidth}
	\centering
	\includegraphics[width=0.5\textwidth]{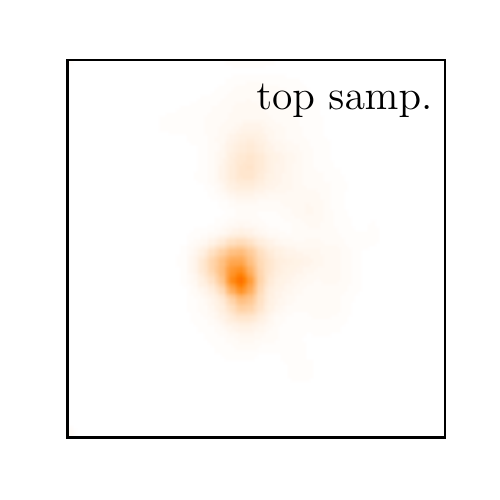}
	\end{subfigure}
	\caption{Average of various jet images. The first two rows account for the direct and
	the inverse tagging problem, and they are showing the average of 10k input and 
	reconstructed images. The last row show an average of 1000 images sampled via
	LMC when training on the two different backgrounds. }
	\label{fig:imgs1}
\end{figure}

The NAE architecture allows us to check explicitly the reconstruction of different jets.
Here, we show the average of two subsamples of QCD and top jets with their reconstruction.
By comparing the input images and the reconstruction we get a better understanding
of the main features learned by the network. These subsamples are shown in 
Fig. \ref{fig:imgs1}. The reconstruction of tops events as signal
shows how the network is only able to reconstruct what's in the training distribution and
therefore ignores additional prongs. In the inverse direction, the network detects all three prongs
while also wrongly reconstructing QCD signal images. In the latter case the main 
contribution to the energy is coming from the intensity of the pixels rather than the 
location of the main prong.

\begin{figure}
	\centering
	\includegraphics[width=0.7\textwidth]{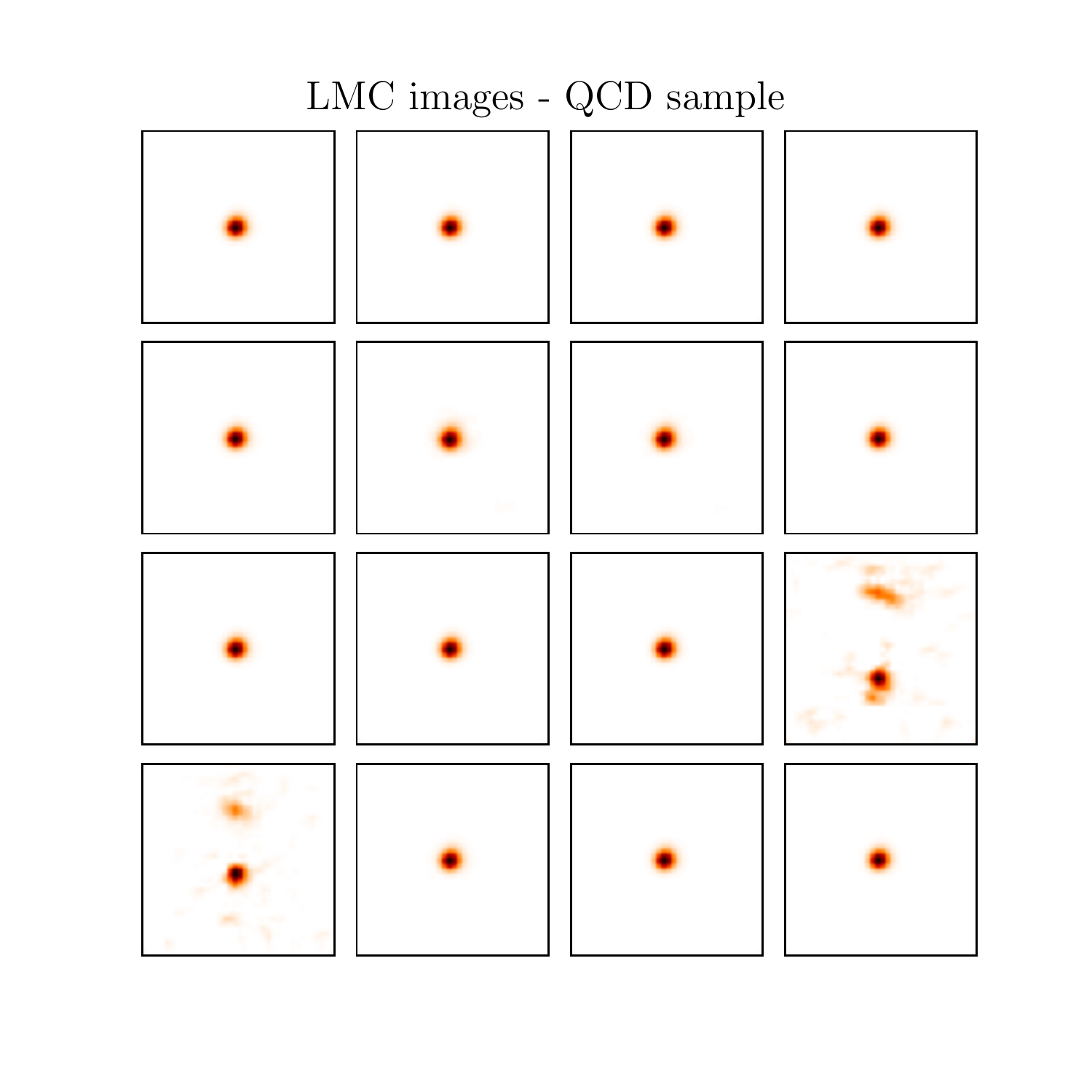} \\
	\includegraphics[width=0.7\textwidth]{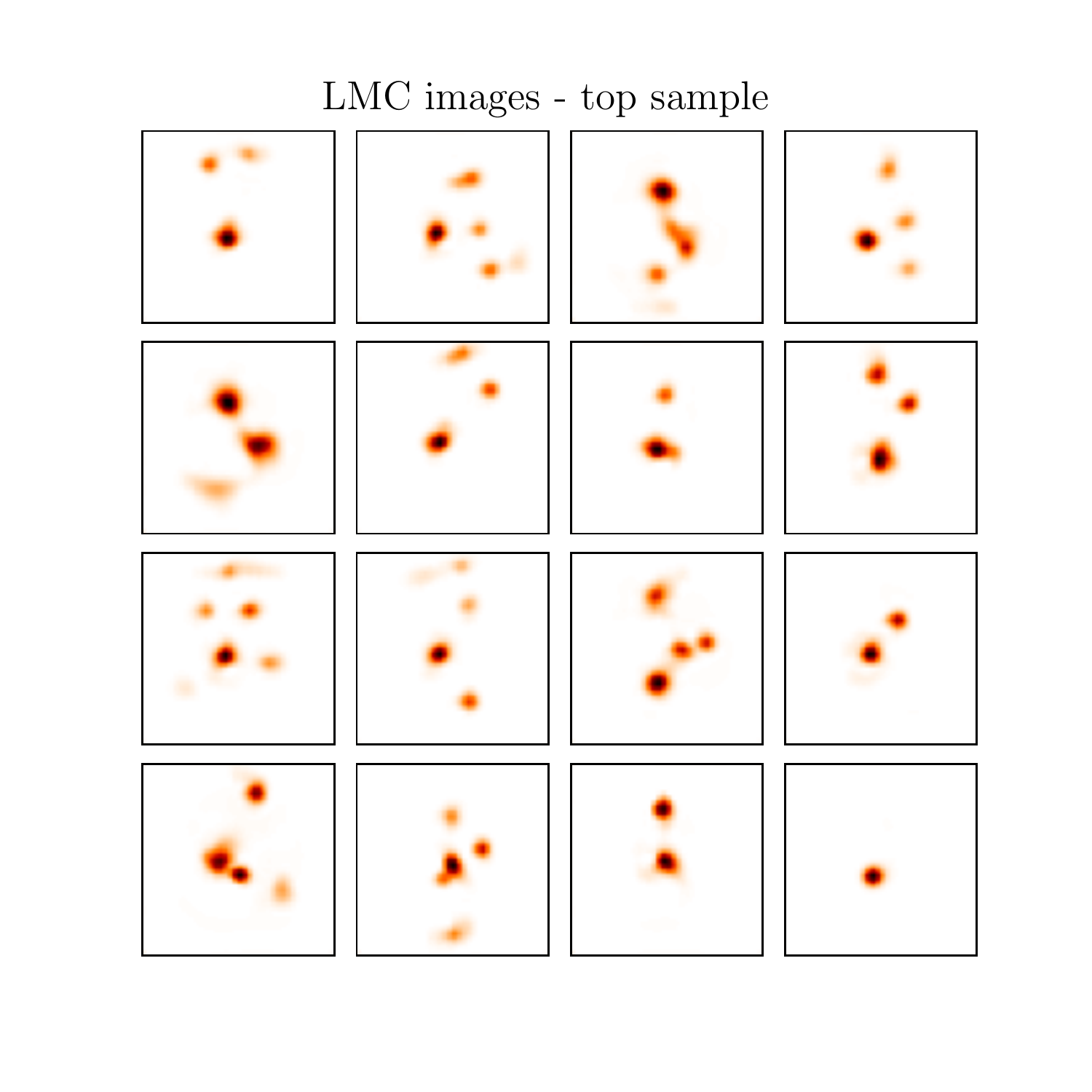}
	\caption{A subset of LMC samples for top (upper) and QCD (lower) tagging.}
	\label{fig:imgs2}
\end{figure}

Furthermore, we can explicitly look at sampled images and compare the average
distribution to the expected one. The last two images of Fig. \ref{fig:imgs1} show
the average sampled distribution for QCD and top tagging. The averaging is performed
on 1000 images sampled after training. In both cases the model distribution has 
converged and resembles the training background. A subset of the LMC samples
is shown in Fig. \ref{fig:imgs2}.

%----------------------------------------------------------
%---------------------------------------------------------
%---------------------------------------------------------
%----------------------------------------------------------

\clearpage

%%%%%%%%%%%%%%%%%%%%%%%%%%%%%%%%%%%%%%%%%%%%%%%%%%%%%%%%%%%%%%%%%
\bibliographystyle{SciPost-bibstyle-arxiv}
\bibliography{main.bib}
\end{document}